\documentclass[12pt,a4paper]{article}
\usepackage{a4,epsf,here}
\def\be{\begin{equation}}
\def\ee{\end{equation}}
\def\bea{\begin{eqnarray}}
\def\eea{\end{eqnarray}}
\def\eq#1{(\ref{#1})}

\def\bs{\bigskip}
\def\ms{\medskip}
\def\fig#1{figure \ref{#1}}
\def\tab#1{table \ref{#1}}
\def\etal{{\it et al}}
\def\bfr{{\bf r}}
\def\d{{\rm d}}

\def\simg{\,\hbox{\kern.1em \lower.6ex \hbox{$\sim$} \kern-1.12em
          \raise.6ex \hbox{$>$} }}
\def\siml{\,\hbox{\kern.1em \lower.6ex \hbox{$\sim$} \kern-1.12em
          \raise.6ex \hbox{$<$} }}
\newcommand{\Figurebb}[9]{
\begin{figure}[H]\begin{center}
\leavevmode
\epsfysize=#7cm
\epsfbox[#2 #3 #4 #5]{#6}
\par
\parbox{#8cm}{
\caption[figure]{\renewcommand{\baselinestretch}{0.8} \small
                                           \hspace{-0.3truecm}#9}
\label{#1}}
\end{center}
\end{figure}
}
\newcommand{\Table}[4]{
\begin{table}[H]\begin{center}{#3}
\parbox{#2cm}{
\vspace{0.5cm}
\caption[table]{\renewcommand{\baselinestretch}{0.8} \small
                                           \hspace{-0.3truecm}#4}
\label{#1}}
\end{center}
\end{table}
}
\begin{document}

\baselineskip 14pt

\centerline{\bf \Large Harmonically trapped fermion gases:}

\bs

\centerline{\bf \Large exact and asymptotic results in arbitrary dimensions}

\bs
\bs
\bs

\centerline{\bf M Brack$^{1,2}$ and M V N Murthy$^{2}$}

\ms
\bs

{\small

\centerline{$^1$Institute for Theoretical Physics, University of
Regensburg, D-93040 Regensburg, Germany}

\centerline{$^2$Institute of Mathematical Sciences, CIT Campus,
                Tharamani, Chennai 600 113, India}

\bs

\centerline{\today}

\bs

\noindent
{\bf \large Abstract}

\ms
\noindent
We investigate the particle and kinetic energy densities of harmonically 
trapped fermion gases at zero temperature in arbitrary dimensions. We
derive analytically a differential equation connecting these densities, 
which so far have been proven only in one or two dimensions, and give
other interesting relations involving several densities or the particle
density alone. We show that in the asymptotic limit of large particle 
numbers, the densities go over into the semi-classical Thomas-Fermi (TF) 
densities. Hereby the Fermi energy to be used in the TF densities is 
identified uniquely. We derive an analytical expansion for the remaining 
oscillating parts and obtain very simple closed forms for the 
leading-order oscillating densities. Finally, we show that the simple 
TF functional relation $\tau_{TF}[\rho]$ between kinetic and particle 
density is fulfilled also for the asymptotic quantum densities $\tau(r)$ 
and $\rho(r)$ including their leading-order oscillating terms. 

\section{Introduction}
\label{secint}

The recent experimental success in confining degenerate Fermi gases in
magnetic traps \cite{jin} has triggered a renewed interest in the
theoretical description of the particle and kinetic energy densities of
harmonically trapped fermions, both at zero
\cite{vig1,glei,bvz,mar1,mar2,vig2,homa} and finite temperatures
\cite{akde,zbsb}. Whereas most of these investigations were limited to
$d=1$, 2 or 3 dimensional systems, general analytical expressions for 
the particle and kinetic energy densities valid for any dimension $d$ 
were given in \cite{bvz,zbsb}. The present paper is devoted to a 
discussion of exact and asymptotic results for particle and kinetic 
energy densities in arbitrary dimensions. We put an emphasis on 
separating their smooth and oscillating parts, identifying analytically 
the smooth parts with those obtained in the semi-classical Thomas-Fermi 
(TF) limit, and finding simple closed expressions for the asymptotically
leading oscillating parts. Using the latter, we find that the TF 
functional relation $\tau_{TF}[\rho]$ between kinetic and particle 
densities holds also for the densities including the leading-order 
quantum oscillations, confirming a recent numerical observation 
\cite{bvz} of this surprising result.

We start by recapitulating some basic definitions given in \cite{bvz}. 
We consider a system of $N$ spin-half fermions, trapped in a spherical 
harmonic potential in $d$ dimensions
\be
V(r) = \frac{m\omega^2}{2}\,r^2\,, \qquad r^2 = x_1^2+x_2^2+\dots+x_d^2\,.
\ee 
We assume that the lowest $M+1$ shells are filled, so that the ground 
state is non-degenerate. The local density then depends only onthe radial 
variable $r$ and is given in terms of the wavefunctions $\phi_i(\bfr)$ by
\be
\rho(r) = 2\!\! \sum_{\epsilon_i\leq E_F} |\phi_i(\bfr)|^2,
\label{rho}
\ee
where the Fermi energy $E_F$ can be identified with the highest filled 
level, $E_F=\epsilon_M=M+d/2$, and a spin degeneracy factor of 2 has been
included. For the kinetic energy density, we consider the two equivalent
expressions
\bea
\tau(r)   & = & - \frac{\hbar^2}{2m}\, 2\!\! \sum_{\epsilon_i\leq E_F}
                   \phi_i^*(\bfr)\nabla^2 \phi_i(\bfr)\,,\label{tau}\\
\tau_1(r) & = & \frac{\hbar^2}{2m}\, 2\!\! \sum_{\epsilon_i\leq E_F}
                   |\nabla\phi_i(\bfr)|^2\,.\label{tau1}
\eea
[Note that in the standard literature on density functional theory, 
$\tau(r)$ sometimes denotes the quantity which we here call $\tau_1(r)$.]
In the presence of time-reversal symmetry (which implies the particle 
number $N$ to be even), they are simply related by
\be
\tau(r) = \tau_1(r) - \frac12\, \frac{\hbar^2}{2m}\, 
             \nabla^2\rho(r)\,.                        \label{taurel}
\ee
A convenient quantity is their average
\be
\xi(r) = \frac12\, [\tau(r)+\tau_1(r)]\,,              \label{xi}
\ee
which is numerically known \cite{bvz,rkb1} to be a smooth function 
whereas $\tau(r)$ and $\tau_1(r)$ have oscillations that are opposite 
in phase (see also \fig{kinf} below). We can then express $\tau(r)$ 
and $\tau_1(r)$ in terms of $\xi(r)$ and $\nabla^2\rho(r)$:
\be
\tau(r) = \xi(r) -\frac14\, \frac{\hbar^2}{2m}\,\nabla^2\rho(r)\,, 
          \qquad\qquad           
\tau_1(r) = \xi(r) +\frac14\, \frac{\hbar^2}{2m}\,\nabla^2\rho(r)\,.
\label{tauxi}
\ee
Eqs.~\eq{rho} -- \eq{tauxi} are exact expressions; for a given number 
of particles or filled shells in any dimension they can be computed 
exactly using the known quantum-mechanical wave functions.

It is well-known numerically that for large numbers of particles, the 
exact densities go over to the smooth TF densities, as also demonstrated
in the two following figures. In \fig{denf} we compare exact 
particle densities $\rho(r)$ with those of the TF model (see section 
\ref{sectf} for their precise definition) for two particular cases in
two and three dimensions. In \fig{kinf} we show a comparison between 
the 

\Figurebb{denf}{10}{50}{795}{390}{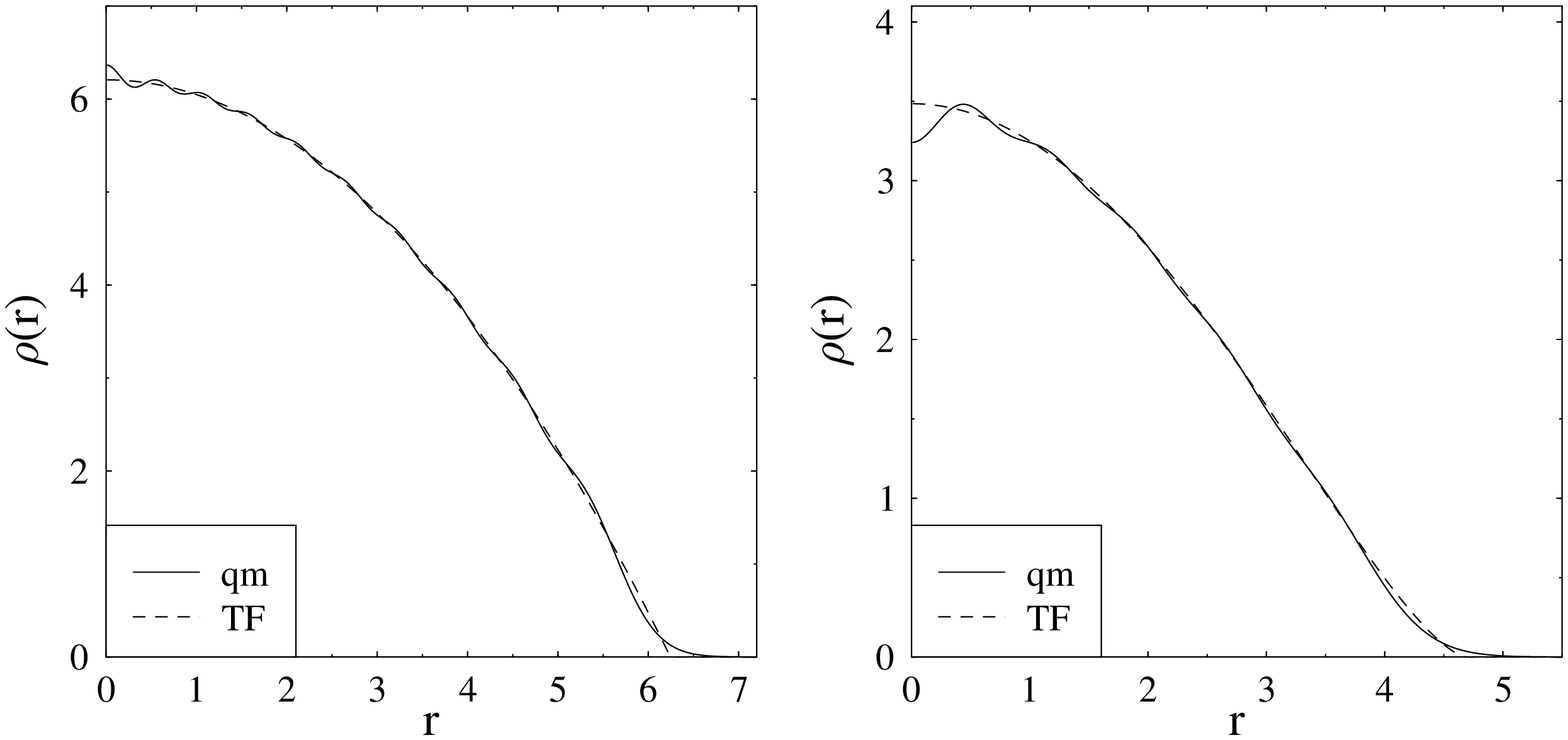}{7}{16}{
Particle densities of fermions in harmonic traps. {\it Left panel:} 
$N=380$ particles filling 19 shells ($M=18$) in 2D; {\it right panel:} 
$N=440$ particles filling 10 shells ($M=9$) in 3D. {\it Solid lines:} 
exact quantum-mechanical densities $\rho(r)$, {\it dashed lines:} 
Thomas-Fermi densities $\rho_{TF}(r)$.
}

\Figurebb{kinf}{10}{50}{795}{390}{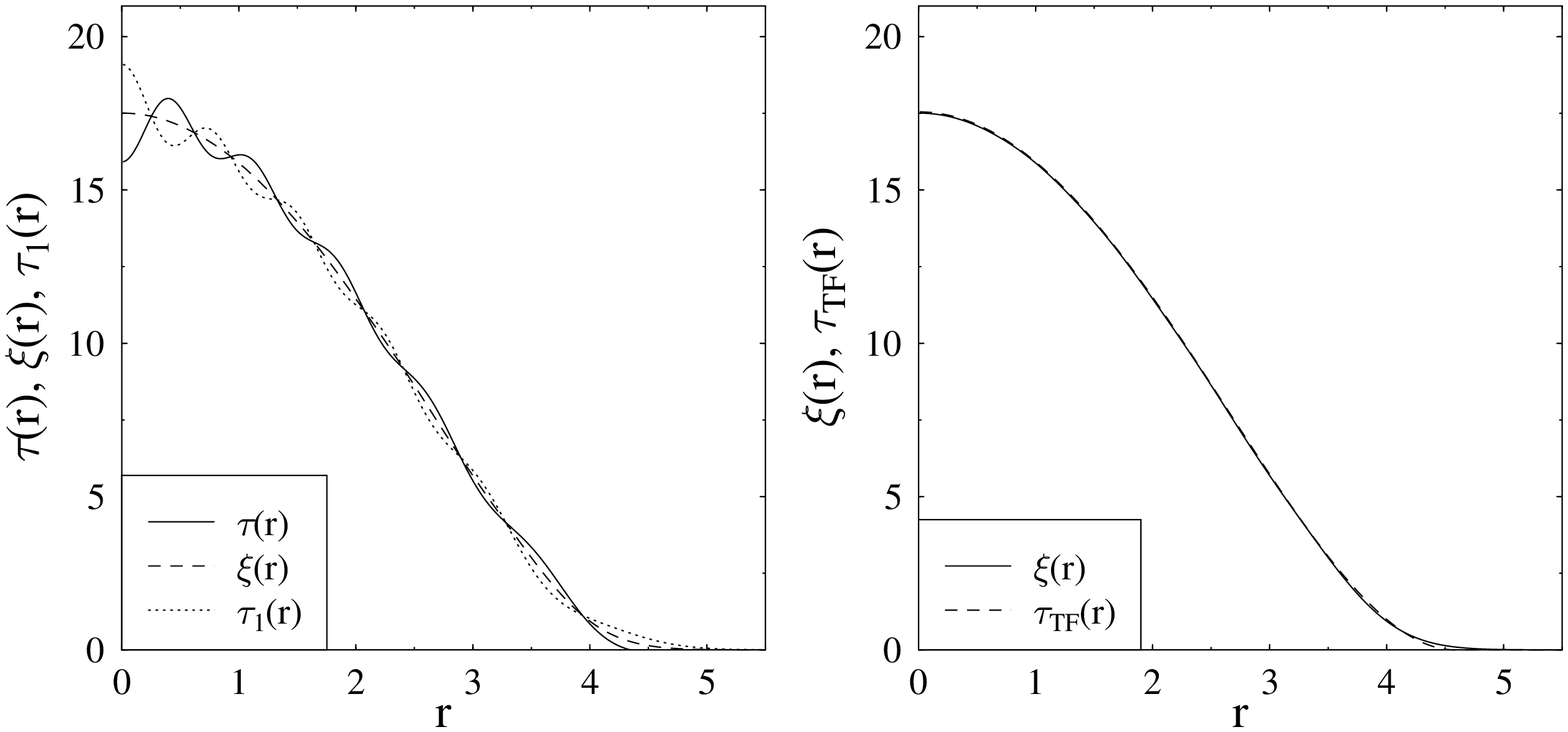}{7}{16}{
Kinetic energy densities of $N=110$ fermions ($M=9$) in a 2D harmonic 
trap. {\it Left panel:} three equivalent quantum-mechanical densities 
integrating to the exact kinetic energy. {\it Right panel:} $\xi(r)$ 
and Thomas-Fermi density $\tau_{TF}(r)$.
}

\noindent
three forms \eq{tau}, \eq{tau1} and \eq{xi} of the kinetic energy 
density, along with that of the TF model. In all cases, the exact 
quantum-mechanical densities oscillate around the TF densities except 
near the turning points where the latter go to zero by definition. In 
particular, the two densities $\tau(r)$ and 
$\tau_1(r)$ have oscillations with exactly opposite phase, which 
renders their average $\xi(r)$ smooth and following very closely the 
TF kinetic energy density. 

This paper is devoted to examining various issues related to the
correspondence between the exact quantum-mechanical densities and their
TF limits, which in the framework of density functional theory correspond
to the local density approximation (LDA). We first derive a homogeneous 
differential equation for the particle density $\rho(r)$, linking it to 
the kinetic energy density $\xi(r)$ in arbitrary dimensions. This has 
been one of the challenges in the density functional theory and has only 
been proved recently \cite{mar1} in $d=2$. An interesting consequence of 
this equation is that the Fermi energy to be used in the TF model is 
given analytically in terms of the number $M$ of the highest filled shell.
 
The surprising success of the TF or LDA formalism in reproducing the 
averaged quantum-mechanical results nearly all the way down to the 
turning point raises an important question about the domain of validity 
of the TF approximation. We examine this question in detail. Using a 
Taylor expansion of the exact densities we show that the leading-order 
terms in the limit of large particle numbers or, equivalently, large
Fermi energies reproduce exactly the TF densities. Furthermore, both  
smooth and oscillating corrections to the TF limits can be gleaned from 
the Taylor expansions. We also use a calculation of the moments of the
densities to buttress our conclusions. While the smooth parts may easily 
be given exactly for even dimensions, they are harder to obtain in odd 
dimensions, necessitating the asymptotic Stirling expansion of Gamma 
functions. Nevertheless, the leading-order results and their 
corrections provide clear pointers to an understanding of the numerical 
results. In particular, we succeed in extracting the asymptotically 
leading-order oscillating terms of the densities in a very simple closed 
analytical form valid for any dimension. The limitations of these
asymptotic terms, and interesting relations between them, are 
investigated numerically.

The remainder part of this paper is organized as follows. In section 
\ref{secdens} we give explicit expressions for the densities using the 
definitions given above. Some of the results are already contained in 
\cite{bvz} but are simplified here and given explicitly, in order to 
make the further analytical calculations more transparent. Section 
\ref{secdifeq} is devoted to the derivation of exact differential 
equations, valid in arbitrary dimensions, that connect the kinetic 
energy and particle densities, and of an equivalent integro-differential 
equation for the particle density that has the form of a 
Schr\"odinger-type eigenvalue equation. In section \ref{sectf} we 
discuss the asymptotic limits of the exact densities for large Fermi
energies $\lambda$ and discuss their precise relation to the TF densities. 
What emerges clearly from this analysis is that the densities can be
uniquely separated into smooth and oscillating parts. The two 
leading-order terms in $\lambda$ of the smooth parts constitute 
analytically the TF densities, whereas the oscillating parts are of 
relative order $1/\lambda$ (or lower), so that they vanish in the 
asymptotic limit. In section \ref{secosc} we discuss the oscillating 
parts of the exact densities. In particular, we show that their 
leading-order contributions, which for any dimension can be given 
analytically in terms of Bessel functions, are eigenfunctions of the 
radial Laplace operator. The TF functional $\tau_{TF}[\rho]$ is shown 
to hold for the asymptotic densities including the leading-order
oscillating terms. The full oscillating parts of the densities can 
be expanded as series of higher-order Bessel functions, which we 
exhibit analytically for even dimensions. Some numerical tests of our 
asymptotic relations are presented in section \ref{secnum}. In section 
\ref{secsum} we finally present a summary of all the new results 
obtained in this paper. Some technical details are given in three
appendices.

\section{Analytical densities for filled spherical shells}
\label{secdens}

Following the general definitions given in the previous section, we quote
here the exact expressions for the densities derived in \cite{bvz} and 
give some new and more general analytical expressions which serve as 
starting points for the following investigations. In the remainder of 
this paper, we work throughout with dimensionless units corresponding to 
the choice $m=\omega=\hbar=1$. The exact quantum-mechanical densities 
$\rho(r)$ and $\xi(r)$ can be written \cite{bvz}, for a given number
$M$ of the last filled shell, in the simple form
\bea
\rho(r,M) & = & \frac{1}{\pi^{\!d/2}}\, 2 \sum_{\mu=0}^M
                F_{M-\mu}^{(d)}(-1)^\mu L_\mu(2x)\, e^{-x}, \label{dden}\\
\xi(r,M)  & = & \frac{1}{\pi^{\!d/2}}\, \frac{d}{2}
                \sum_{\mu=0}^M G_{M-\mu}^{(d)}(-1)^\mu L_\mu(2x)\, e^{-x},
\label{dkin}
\eea
where $x=r^2$ and $L_\mu(2x)$ are Laguerre polynomials. The coefficients 
$F_\nu^{(d)}$, $G_\nu^{(d)}$ are given by
\be
F_\nu^{(d)} = \nu + 1 + \sum_{m=1}^{[\nu/2]} (\nu+1-2m)\,g_m^{(d)}\,,\qquad
G_\nu^{(d)} = (\nu + 1)^2 + \sum_{m=1}^{[\nu/2]} (\nu+1-2m)^2\,g_m^{(d)}\,, 
\label{FG}
\ee
where $[\nu/2]\equiv{\rm integer}\,(\nu/2)$. The factors $g_m^{(d)}$  
are defined through 
\be
(1-x)^{(1-d/2)} = 1 + \sum_{m=1}^\infty g_m^{(d)} x^m \qquad
\Longrightarrow \qquad
g_m^{(d)} = \frac{1}{m!}\frac{\Gamma(d/2+m-1)}{\Gamma(d/2-1)}\,.
\ee
The results are particularly simple for $d=2$, where all the 
$g_m^{(2)}$ are identically zero:
\be
F_\nu^{(2)} = \nu + 1\,, \qquad G_\nu^{(2)} = (\nu + 1)^2\,. \label{FG2} 
\ee

In \cite{bvz} it was stated that $F^{(d)}_{\nu}$ and $G^{(d)}_{\nu}$ 
can be given in closed analytical form only for even dimensions. We 
now have found \cite{mapl} that this actually can be done in any 
dimension $d$, albeit separately for even and odd indices: 
\be
F_{2n}^{(d)} = (d/2+2n)\,\frac{\Gamma(d/2+n)}{n!\,\Gamma(d/2+1)}\,,\qquad
F_{2n+1}^{(d)} = 2\,\frac{\Gamma(d/2+n+1)}{n!\,\Gamma(d/2+1)}\,, \label{Fn}
\ee
\be                                                    
G_{2n}^{(d)} = \frac14\,(32\,n^2+d^2+16\,nd+2d)\,
                   \frac{\Gamma(d/2+n)}{n!\,\Gamma(d/2+2)}\,, \qquad
G_{2n+1}^{(d)} = 2\,(4n+d+2)\,\frac{\Gamma(d/2+n+1)}{n!\,
                                           \Gamma(d/2+2)}\,.     \label{Gn} 
\ee
From \eq{Fn} one finds that the $F_\nu^{(d)}$ obey the following summation 
relations: 
\be
2\sum_{\nu=0}^M (-1)^{M-\nu} F_\nu^{(d)} = F_{M+1}^{(d)} \quad
                                        (M\;{\rm even})\,, \qquad
2\sum_{\nu=0}^M (-1)^{M-\nu} F_\nu^{(d)} = F_M^{(d)} \quad
                                        (M\;{\rm odd})\,. \label{Fodsum}
\ee

Using the above formulae, we can readily give closed expressions for the 
densities at $r=0$. For even and odd $M$, respectively, we get for the 
particle densities
\bea
\rho_{\rm even}(0,M) & = & \!\frac{2}{\pi^{d/2}}\,\frac{\Gamma(M/2+d/2+1)}
                           {\Gamma(d/2+1)\,\Gamma(M/2+1)}\quad\;
                       = \;\frac{2}{\pi^{d/2}}\,\frac{(M+d)!!}{d!!\,M!!}
                           \qquad \;\;\quad(M\;{\rm even})\,,\nonumber\\
\rho_{\rm odd}(0,M)  & = & \!\frac{2}{\pi^{d/2}}\,\frac{\Gamma(M/2+d/2+1/2)}
                           {\Gamma(d/2+1)\,\Gamma(M/2+1/2)}\;
                   =\;\frac{2}{\pi^{d/2}}\,\frac{(M-1+d)!!}{d!!\,(M-1)!!}
                           \qquad (M\;{\rm odd})\,,        \label{rhozero}
\eea
and for the kinetic energy densities
\bea
\xi_{\rm even}(0,M) & = &  \!\frac{(4M+d+2)}{(d+2)}\,\,\rho_{\rm even}(0,M)\,,
                           \nonumber\\
\xi_{\rm odd}(0,M) & = & \!\frac{(4M+3\,d+2)}{(d+2)}\,\rho_{\rm odd}(0,M)\,.
                                                            \label{xizero}
\eea
On the right-hand sides in \eq{rhozero}, which are easier to evaluate 
manually, we have used the double factorials defined (for integer $n$) by
\bea
 (2n)!!   & = & 2n\,(2n-2)\cdot\cdot\cdot 2
            =   \sqrt{\pi}\,\Gamma(2n+1)/2^n\Gamma(n+1/2)\,,
                \qquad 0\,!! = 1\,,\nonumber\\
 (2n-1)!! & = & (2n-1)(2n-3)\cdot\cdot\cdot1
            =   2^n\,\Gamma(n+1/2)/\!\sqrt{\pi}\,,\qquad\;\; (-1)!! = 1\,.
\eea
Note that the densities \eq{rhozero} for any even $M$ and the next odd
$M+1$ are identical. This is due to the alternating parities of the 
quantum states in successive shells. While filling a shell with odd 
parity (ie odd $M+1$), no contribution is added to the density 
$\rho_{\rm even}(r,M)$ at $r=0$. For even dimensions $d$, the 
expressions \eq{rhozero} and \eq{xizero} reduce to polynomials of order 
$d/2$ and $d/2+1$ in $M$, respectively. For odd $d$, they can be 
simplified only in the large-$M$ limit discussed in section \ref{sectf}.

For later reference, we give here also the total particle number $N$ and 
energy $E$ as functions of the shell number $M$, making use of the fact
that the degeneracy of the $m$-th level $\epsilon_m=m+d/2$ is given by
the binomial coefficient ${m+d-1\choose d-1}$:
\bea 
N(M) & = & 2\,\sum_{m=0}^M {m+d-1\choose d-1} = 2\, {M+d\choose M}
                             = 2\,\frac{(M+d)!}{d!\,M!}\,, \label{NofM}\\
E(M) & = & 2\,\sum_{m=0}^M {m+d-1\choose d-1}\left(m+d/2\right)
                 = d\,(2M+d+1)\,\frac{(M+d)!}{(d+1)!\,M!}\,. \label{EofM}
\eea

\newpage

\section{Exact differential equations for densities in arbitrary dimensions}
\label{secdifeq}

As stated by Minguzzi \etal~\cite{mar1}, one of the challenges in density
functional theory is to be able to directly calculate the particle density,
given the potential energy, without recourse to solving the Schr\"odinger
equation for the wave functions. They have achieved this aim for harmonic
confinement of independent fermions in $d=2$. The central result of their
paper \cite{mar1} is the following differential equation for the density
\be
\xi(r) = \frac{1}{2}\left[\frac18\,\Delta\rho(r)
         +\rho(r)\left(M+\frac{3}{2}-\frac12\,r^2\right)\right],
\label{difeq2d}
\ee
where $\Delta$ is the Laplacian operator $\Delta=\nabla^2$ in two 
dimensions.

Indeed, we now prove that such a differential equation may be derived in 
arbitrary dimension $d$. Furthermore, using a ``differential virial 
theorem'' derived recently \cite{zbsb} for any $d$, we show that a 
linear, homogeneous third-order differential equation for the particle 
density may be established in arbitrary dimensions, like it has been
done \cite{mar1} in the case of $d=2$.

In order to proceed, we note that the radial part of the Laplacian 
in $d$ dimensions is given by:  
\be
\Delta = \frac{\d^2}{\d r^2}+\frac{(d-1)}{r}\frac{\d}{\d r}\,.
\ee
For the following  derivation it is useful to express it in terms of 
the variable $x=r^2$:
\be
\Delta = 4\,x\,\frac{\d^2}{\d x^2}+2\,d\,\frac{\d}{\d x}\,.
\ee
The Laplacian acting on the particle density (\ref{dden}) is given by 
\be
\Delta\rho(x) =  \frac{1}{\pi^{\!d/2}}\, 2 \sum_{\mu=0}^M
                 F_{M-\mu}^{(d)}(-1)^\mu 
                 [(4x-2d-8\mu)L_\mu(2x)+(4d-8)L'_\mu(2x)]\, e^{-x}, 
                                                         \label{Lden}
\ee
where we have made use of the differential equation for the Laguerre 
polynomials \cite{abro}
\be 
y\,L''_\mu(y) + (1-y)L'_\mu(y) + \mu \,L_\mu(y)=0\,,
\ee
and the primes denote derivatives with respect to the full argument of 
a function. Manipulating the sums in (\ref{Lden}), we obtain
\be
\frac{1}{8}\,\Delta\rho(x)+(M+1+d/4-x/2)\,\rho(x) = 
                        \frac{4}{d}\,\xi(x)+ {\cal R}(x)\,, \label{rdiff}
\ee
where ${\cal R}(x)$ is given by
\be
{\cal R}(x) = \frac{2}{\pi^{\!d/2}}\sum_{\mu=0}^M (-1)^\mu
         \left\{\left[(M+1-\mu)\,F_{M-\mu}^{(d)}-G_{M-\mu}^{(d)}\right] 
         L_\mu(2x)+(d/2-1)\,F_{M-\mu}^{(d)} L'_\mu(2x) 
         \right\} e^{-x}.                               \label{residue}
\ee
Note that ${\cal R}(x)$ is identically zero for $d=2$, due to \eq{FG2}, 
so that \eq{rdiff} reduces readily to \eq{difeq2d}. For arbitrary 
dimensions $d$, now, ${\cal R}(x)$ can be cast into the following form
(see appendix \ref{appdif} for details): 
\be
{\cal R}(x) = \frac{(d-2)}{d}\,\xi(x) - \frac{(d-2)}{4}\,\rho(x)\,.
\ee
Substituting this into \eq{rdiff}, we arrive at the differential equation
\be
\xi(r) = \frac{d}{(d+2)}\left[\frac18\,\Delta\rho(r)
         +\rho(r)\left(\lambda_M-\frac12\,r^2\right)\right],
\label{difeq}
\ee
which is valid for any $d$ and obviously reduces to \eq{difeq2d} for 
$d=2$ as given in \cite{mar1}. Here $\lambda_M$ is given by  
\be
\lambda_M=M+(d+1)/2\,,
\label{lambda}
\ee
which corresponds to the average of the highest filled and the lowest 
unfilled level and will be identified as the Fermi energy of the smooth 
densities in the large-$M$ limit (see section \ref{sectf} below).

Having established equation (\ref{difeq}), we now proceed to derive
equations for the particle density alone by eliminating the kinetic 
energy density $\xi(r)$. To this end we use an integral equation which 
has recently been derived in \cite{zbsb} for arbitrary dimensions $d$, 
and its equivalent differential form:
\be
\xi(r) = \frac{d}{2}\,\int_r^\infty r'\rho(r')\,\d r' \quad
\Longleftrightarrow \quad
\frac{\d}{\d r}\,\xi(r) = -\frac{d}{2}\,r\rho(r)\quad
\Longleftrightarrow \quad
\frac{\d}{\d x}\,\xi(x) = -\frac{d}{4}\,\rho(x)\,.
\label{inteq}
\ee
The differential form had earlier been derived for $d=2$ in \cite{mar1} 
and termed ``differential virial theorem''. Its version for $d=3$ had 
been correctly guessed in \cite{mar2}. Inserting the integral form of 
$\xi(r)$ into \eq{difeq}, we obtain the following integro-differential 
equation for $\rho(r)$ with eigenvalue $\lambda_M$
\be
-\frac18\,\Delta\rho(r) + \frac12\,r^2\rho(r) 
+ \frac{(d+2)}{2}\int_r^\infty r'\rho(r')\,\d r'
=\lambda_M\,\rho(r)\,.
\label{schreq}
\ee
This has the form of a Schr\"odinger equation for $\rho(r)$ with an
additional non-local potential. Differentiating both sides of 
\eq{schreq}, we can write it as a third-order differential equation 
for $\rho(r)$:
\be
\frac18\,\frac{\d}{\d r}\,\Delta\rho(r)
+\left(\lambda_M-\frac12\,r^2\right)\!\frac{\d}{\d r}\,\rho(r)
+\frac{d}{2}\,r\,\rho(r) = 0\,.
\ee
This equation has been previously derived for $d=1$ in \cite{lama} 
and for $d=2$ in \cite{mar1}. Its form for $d=3$ has been surmised and 
numerically tested in \cite{mar2}, and general solutions for $\rho(r)$ 
in the case $d=3$ were discussed in \cite{homa}. 

Using the relation \eq{tauxi}, we may eliminate the Laplacian term
in \eq{difeq} in favour of the kinetic energy density $\tau(r)$ to 
obtain the following exact relation between three densities
\be
\tau(r) = \rho(r)\left(\lambda_M-\frac12\,r^2\right)
                         -\frac{2}{d}\,\xi(r)\,,         \label{taueq}
\ee
which will prove useful in the discussion of section \ref{secosc}.

Integrating the differential equation \eq{difeq} over the whole space,
using the measure
\be
\int \d^d r\,(\dots) = \frac{\pi^{d/2}}{\Gamma(d/2)}
                     \int_0^\infty x^{d/2-1}\d x\,(\dots)\,,
\label{volel}
\ee
and exploiting the virial theorem, we find
\be
\lambda_M N(M) = \frac{d+1}{d}\,E(M)\,,
\ee
which using \eq{NofM} and \eq{EofM} unambiguously confirms the choice 
of the Fermi energy \eq{lambda}.

\newpage

\section{Large-$M$ limits and relation to the Thomas-Fermi densities}
\label{sectf}

We next address the question as to how the well-known Thomas-Fermi (TF) 
results are obtained from the exact densities \eq{dden}, \eq{dkin} in 
the limit of large particle numbers or, equivalently, in the large-$M$ 
limit. The TF densities are given by (see eg chapter 4 
in \cite{book})
\bea
\rho_{TF}(r,\lambda) & = & \frac{1}{(2\pi)^{d/2}}\,\frac{4}{d}\,
                     \frac{1}{\Gamma(\frac{d}{2})}\,
                     (\lambda-r^2\!/2)^{d/2}\,\Theta(\lambda-r^2\!/2)\,,
\label{tfden}\\
\xi_{TF}(r,\lambda) = \tau_{TF}(r,\lambda) 
                     & = & \frac{1}{(2\pi)^{d/2}}\,\frac{4}{(d\!+\!2)}\,
                     \frac{1}{\Gamma(\frac{d}{2})}\,
                     \!(\lambda-r^2\!/2)^{d/2+1}\,\Theta(\lambda-r^2\!/2)\,,
\label{tfkin}
\eea
where $\lambda$ is a Fermi energy that is usually determined by
normalizing $\rho_{TF}(r,\lambda)$ to the exact particle number $N$. 
The Heavyside step function $\Theta(\lambda-r^2\!/2)$ ensures that these 
densities are identically zero outside the classical turning point 
$r_0=\sqrt{2\lambda}$. 

The proof that the above TF densities can be obtained from the exact 
quantum-mechanical densities in the large-$M$ limit is by no means 
trivial, since the summations in \eq{dden} and \eq{dkin} cannot be done 
analytically in closed form. To our knowledge, a rigorous proof for 
$d=2$ has been given only very recently in \cite{zbsb}, whereby 
$\lambda$ was found to be $\lambda=M+3/2$. 
In \cite{glei} it was shown for $d=1$ that the exact density at 
$r=0$ is asymptotically given by $\rho_{TF}(0)$ for large particle 
numbers, and an expression for the leading-order oscillating part 
$\delta\rho(x)=\rho(x)-\rho_{TF}(x)$, valid for small enough $x$, 
was given.

We first observe that the differential equation (\ref{difeq}) is also 
satisfied by the corresponding TF densities in the limit of large $M$, 
and hence of large $\lambda_M$, since the Laplacian acting on the TF 
density produces terms of lower order in $\lambda_M$. In the following 
we shall show, indeed, that for any dimension $d$ the exact densities 
given in \eq{dden} and \eq{dkin} in the large-$M$ limit go over into 
the TF functions \eq{tfden} and \eq{tfkin}, plus smooth corrections of 
order $1/M^2$ relative to the leading-order TF terms, plus oscillating
terms that are of relative order $1/M$ or lower and will be discussed
in section \ref{secosc}. Hereby it is essential that the Fermi energy 
$\lambda$ be identified with the $\lambda_M$ given in \eq{lambda}. 
Whereas the proof given in \cite{zbsb} for $d=2$ can easily be 
generalized to any even $d$ (see section \ref{secosc} and appendix 
\ref{apposc}), the cases with odd $d$ turn out to be more difficult to 
handle. We will first discuss the separation of smooth and oscillating
parts in general. In the two ensuing subsections, we shall then use 
Taylor expansions of the exact densities and a calculation of their 
moments in order to prove our claims. 

\subsection{Separation of smooth and oscillating terms}

For the following discussions, it is useful to separate the densities 
into average and oscillating parts:
\be
\rho(r) = {\bar \rho}(r) + \delta\rho(r)\,,\qquad
\xi(r)  = {\bar \xi}(r)  + \delta\xi(r)\,.~~~~~~
\label{densep}
\ee
This can be achieved using the following trick. We define the smooth 
parts by
\bea
{\bar\rho}(r,M) & = & \frac12\,[\rho_{\rm even}(r,M)+\rho_{\rm odd}(r,M)]\,,
\qquad\nonumber\\ 
{\bar\xi}(r,M)  & = & \frac12\,[\xi_{\rm even}(r,M)+\xi_{\rm odd}(r,M)]\,,
\label{avden} 
\eea
whereby the formal expressions obtained from \eq{dden} and \eq{dkin} 
for even and odd $M$ must be taken at the same fixed value of $M$ and 
inserted on the right-hand sides above, irrespectively of $M$ being odd 
or even. In the same fashion, the oscillating parts are defined by 
\bea
\delta\rho(r,M) & = & (-1)^M\frac12\,
                      [\rho_{\rm even}(r,M)-\rho_{\rm odd}(r,M)]\,,\qquad
                      \nonumber\\
\delta\xi(r,M)  & = & (-1)^M \frac12\,
                      [\xi_{\rm even}(r,M)-\xi_{\rm odd}(r,M)]\,.
\label{oscden}
\eea
The resulting expressions are then valid for either even or odd $M$. The 
sign $(-1)^M$ in front of the oscillating terms is easy to understand
due to the alternating parities of successive oscillator shells, as 
discussed in section \ref{secdens} after the equations \eq{rhozero} for 
the density at $r=0$. The alternating sign of $\delta\rho(0,M)$ for 
successive shell numbers $M$ is intimately connected with the oscillating
behaviour of $\delta\rho(r,M)$ as a function of $r$. For even $M$, it 
must have a maximum at $r=0$ since an even shell always contains an
$s$ (zero angular momentum) state. After filling the next (odd) shell, 
it has acquired a minimum, and so on. This is also illustrated in the 
numerical examples shown in \fig{denf}.

From the explicit results given below it can be verified
that inserting \eq{densep} into \eq{difeq}, the smooth parts cancel 
identically and the following differential equation holds for the 
oscillating parts alone:
\be
\delta\xi(r) = \frac{d}{(d+2)}\left[\frac18\,\Delta\delta\rho(r)
         +\delta\rho(r)\left(\lambda_M-\frac12\,r^2\right)\right].
\label{difeqosc}
\ee
Similarly, all smooth terms cancel also from \eq{taueq}, so that we 
have the equation:
\be
\delta\tau(r) = \delta\rho(r)\,\left(\lambda_M-\frac12\,r^2\right)
                -\frac{2}{d}\,\delta\xi(r)\,.
\label{taueqosc}
\ee
We finally note that the differential equation given in \eq{inteq} for 
the full densities holds separately for their smooth and oscillating 
parts as well as for the TF densities alone:
\be
\frac{\d}{\d r}\,{\bar\xi}(r) = -\frac{d}{2}\,r\,{\bar\rho}(r)\,,\qquad
\frac{\d}{\d r}\,\delta\xi(r) = -\frac{d}{2}\,r\,\delta\rho(r)\,,\qquad
\frac{\d}{\d r}\,\xi_{TF}(r)  = -\frac{d}{2}\,r\,\rho_{TF}(r)\,.
\label{difav}
\ee

Let us now look at the densities at the centre of the system, $r=0$.
As stated at the end of section \ref{secdens}, for even dimensions $d$ 
the expressions \eq{rhozero} and \eq{xizero} for $\rho(0)$ and $\xi(0)$
are polynomials in $M$ of order $d/2$. It is easy to see that the 
coefficients of their highest power in $M$ are identical with those of 
$\rho_{TF}(0,\lambda_M)$ and $\xi_{TF}(0,\lambda_M)$, respectively, so 
that the latter become exact in the large-$M$ limit. 
For odd $d$, we have to use the Stirling expansion 
of the Gamma functions in \eq{rhozero}, valid for large $M$ (cf 
\cite{abro}, eq 6.1.37), in order to find the same result. It is even 
more instructive to look separately at the smooth and oscillating parts.
We find that in the large-$M$ limit, the smooth parts reproduce the 
{\it two} highest powers of $M$ in the corresponding TF densities at 
$r=0$ correctly:
\bea
{\bar\rho}(0,M) & = & \frac{1}{(2\pi)^{d/2}}\,\frac{4}{d}\,
                      \frac{1}{\Gamma(\frac{d}{2})}\,\lambda_M^{d/2}
                      \left[1+{\cal O}(1/M^2)+\dots\right],\nonumber\\
{\bar\xi}(0,M)  & = & \frac{1}{(2\pi)^{d/2}}\,\frac{4}{(d+2)}\,
                      \frac{1}{\Gamma(\frac{d}{2})}\,\lambda_M^{d/2+1}
                      \left[1+{\cal O}(1/M^2)+\dots\right],\label{denavas}
\eea
whereby the series in $1/M$ stop at order $M^0$ for even $d$. The 
oscillating parts are found to be
\bea
\delta\rho(0,M) & = & (-1)^M\frac{1}{(2\pi)^{d/2}}\,
                      \frac{1}{\Gamma(\frac{d}{2})}\,\lambda_M^{d/2-1}
                      \left[1+{\cal O}(1/M)+\dots\right],\nonumber\\  
\delta\xi(0,M) & = & (-1)^M\frac{1}{(2\pi)^{d/2}}\,
                      \frac{d(d-2)}{16\Gamma(\frac{d}{2})}\,\lambda_M^{d/2-2}
                      \left[1+{\cal O}(1/M)+\dots\right].\label{denoscas}
\eea
We see that $\delta\rho(0)$ is of order $1/\lambda_M$ lower than the
leading power in the smooth density, whereas the suppression factor
in the  $\delta\xi(0)$ is $1/\lambda_M^3$. This hints at the numerically
known fact that $\xi(r)$ is much smoother than $\rho(r)$.

\subsection{Taylor expansion of the densities}

We now want to study how the above results extend to $x>0$.
In the appendix \ref{apptay}, we give the explicit Taylor expansions
of the exact densities \eq{dden} and \eq{dkin} in powers of $x=r^2$:
\be
\rho^{(d)}(x,M) = \sum_{m=0}^\infty \rho^{(m)}(d,M)\,x^m\,,\qquad\qquad
\xi^{(d)}(x,M)  = \sum_{m=0}^\infty \xi^{(m)}(d,M)\,x^m\,.
\label{tayden}
\ee
The expansion coefficients cannot be given in closed form for any $d$,
but must be evaluated separately for even and odd values of $M$. From
these, one can reconstruct separately the expansion coefficients of the 
smooth and oscillating parts according to \eq{avden} and \eq{oscden}. 

As shown in detail in appendix \ref{apptay}, we find the following
structure of the results for even dimensions $d$. The smooth parts are
finite polynomials in $x$ which have exactly the structure of the TF 
densities \eq{tfden} and \eq{tfkin}, including the step functions that
cut these densities at the classical turning point. However, only the 
two highest powers of $M$ appearing in the coefficients of each power 
of $x$ agree with those in the corresponding powers of the Fermi energy 
$\lambda_M=M+(d+1)/2$ \eq{lambda} appearing in the TF densities. In
other words, the smooth parts go in the large-$M$ limit over into the 
TF densities plus smooth terms of order $1/M^2$ and lower relative to 
the leading TF terms:
\bea
{\bar\rho}(r,M) & = & \rho_{TF}(r,\lambda_M)\,
                      \left[1+{\cal O}(M^{-2})+\dots\right],\nonumber\\
{\bar \xi}(r,M) & = & \xi_{TF}(r,\lambda_M)\,
                      \left[1+{\cal O}(M^{-2})+\dots\right],\label{TFas}
\eea
which is the straightforward extension of \eq{denavas}. To put this 
structure into evidence, we list here the complete smooth densities for 
the lowest even dimensions, re-expressed in terms of the Fermi energy. 
For $d=2$ we get, with $\lambda=M+3/2$:
\bea
{\bar\rho}^{(2)}(r) & = & \frac{1}{\pi}\,(\lambda-r^2\!/2)\,
                          \Theta(\lambda-r^2\!/2)\,,\nonumber\\
{\bar\xi}^{(2)}(r)  & = & \frac{1}{2\pi}\left[(\lambda-r^2\!/2)^2
                          -1/4\right]\Theta(\lambda-r^2\!/2)\,.\label{avd2}
\eea
For $d=4$ we get, with $\lambda=M+5/2$:
\bea
{\bar\rho}^{(4)}(r) & = & \frac{1}{4\pi^2}\left[(\lambda-r^2\!/2)^2
                          -3/4\right]\Theta(\lambda-r^2\!/2)\,,\nonumber\\
{\bar\xi}^{(4)}(r)  & = & \frac{1}{6\pi^2}\left[(\lambda-r^2\!/2)^3
                          -7\lambda/4+9\,r^2\!/8\right]
                           \Theta(\lambda-r^2\!/2)\,.\label{avd4}
\eea

The oscillating parts, characterized by the alternating sign $(-1)^M$, 
contain infinite power series in $x$ that can be resummed into oscillating 
functions of $x$, as shown in appendix \ref{apposc} and further discussed
in section \ref{secosc}. After resummation, they are found to be of order 
$1/M$ or lower, relative to the leading-order terms of the smooth parts, 
so that they become negligible in the large-$M$ limit.

The situation is more complicated for odd dimensions $d$. This becomes
evident when we compare the powers of $\pi$ appearing in the denominators
of the exact densities \eq{dden}, \eq{dkin} and their TF expressions
\eq{tfden}, \eq{tfkin}: whereas they are the same for even $d$, they
are different by a factor $\sqrt{\pi}$ for odd $d$. The reason is that 
in order to obtain the TF densities in the large-$M$ limit, one has to 
perform a large-$M$ Stirling expansion of the $\Gamma$ functions in 
\eq{rhozero}, as seen above for $\rho(0)$.
In doing so for specific odd values of $d$, we find that the resulting
densities have the same structure as for even $d$. Their smooth parts 
yield the Taylor expansions of the TF densities \eq{tfden}, which here
are infinite due to the half-integer power $d/2$:
\be
{\bar\rho}^{(d)}(r,M)\,=\,\rho_{TF}^{(d)}(0,\lambda)
                          \left[1-\frac{d}{2}\,\frac{x}{2\lambda}
                          +\frac{d(d-2)}{8}\,\left(\frac{x}{2\lambda}\right)^2
                          -\dots \right]\left[1+{\cal O}(M^{-2})+\dots\right],
\ee
with $\lambda=\lambda_M$ given in \eq{lambda}, and analogously for the 
average kinetic energy densities ${\bar\xi}^{(d)}(r,M)$. The oscillating 
parts are again at least of order $1/M$ relative to the leading smooth 
parts and therefore vanish in the large-$M$ limit. 

Summarizing up to this point, we find that the densities $\rho(r)$ and 
$\xi(r)$ can uniquely be decomposed into smooth and oscillating parts. 
In the large-$M$ limit, the smooth parts ${\bar\rho}(r)$ and
${\bar\xi}(r)$ reach their TF forms \eq{tfden}, 
\eq{tfkin} in terms of the Fermi energy $\lambda_M$ given in 
\eq{lambda}, plus terms of relative order $1/M^2$ and lower at each 
power of $x$ in their Taylor expansions. The oscillating parts 
$\delta\rho(r)$ and $\delta\xi(r)$ are of order $1/M$ and $1/M^3$, 
respectively, compared to the leading-order TF terms, as discussed
explicitly in section \ref{secosc}. This clearly establishes the
TF densities as the asymptotic large-$M$ limits of the exact 
quantum-mechanical densities. 

\subsection{Calculation of moments of the densities}

As an alternative way to demonstrate the asymptotic equality of the
exact and TF densities, we will now calculate their moments and show 
that they are all identical to the two leading orders of $M$. Since the 
densities are functions only of $x=r^2$, it is sufficient to compute 
only the even moments of $r$, i.e., the moments of the variable $x$. 
We define them as
\be
C^{(m)} = \int \d^dr \,x^{m}\,\rho(x)\,, \qquad
D^{(m)} = \int \d^dr \,x^{m}\,\xi(x)\,.                   \label{moms}
\ee
Integrating $D^{(m)}$ by parts using \eq{inteq} and \eq{volel}, one 
easily sees that the following relation holds:
\be
D^{(m)} = \frac{d}{(4m+2d)}\,C^{(m+1)}\,.                \label{virial}   
\ee
Note that this relation for $m=0$ yields the virial theorem, since
$D^{(0)}$ is the kinetic energy and $C^{(1)}$ twice the potential
energy. The moments of the TF density are readily found to be
\be
C_{TF}^{(m)} = \int \d^dr \,x^{m}\rho_{TF}(x,\lambda_M)
             = \frac{\Gamma(d/2+m)}{\Gamma(d/2)}\,
               \frac{2^{m+1}}{(m+d)!}\,\lambda_M^{(m+d)}.  \label{momtf}
\ee
The corresponding moments of the exact density might be obtained by 
direct integration of $x^m$ over the expression \eq{dden} which leads to
\be
C_{qm}^{(m)} = \frac{\Gamma(d/2+m)}{\Gamma(d/2)}\,2
               \sum_{\mu=0}^{M} F_{M-\mu}^{(d)}(-1)^{\mu} 
               F(-\mu,m+d/2;1;2)\,,                        \label{momdqm} 
\ee
where $F(a,b;c;z)$ is a hypergeometric function. However, we were not able 
to do the above sums analytically for arbitrary $d$. We therefore turn to
the case $d=1$, where the moments are more easily obtained from a 
summation of the corresponding matrix elements. The matrix element of 
$x^m$ in the $n$-th quantum state of the one-dimensional harmonic
oscillator is found to be
\be
\langle n|x^m|n\rangle = \frac{(2m)!}{2^{2m}m!}\,(-1)^nF(-n,m+1;1;2)\,.
\ee
The exact moments are therefore
\be
C_{qm}^{(m)} = \frac{(2m)!}{2^{2m}m!}\,2\sum_{n=0}^M(-1)^nF(-n,m+1;1;2)\,.
\label{momqm}
\ee
The above hypergeometric functions are easy to create recursively
by the relation
\be
(m+1)\,{\cal F}_{m+1}(n) = (2n+1)\,{\cal F}_m(n)+m\,{\cal F}_{m-1}(n)\,,
\ee
where
\be
{\cal F}_m(n) = (-1)^nF(-n,m+1;1;2)\,,
\ee
and using $F(-n,1;1;2)=(-1)^n$ (see \cite{grry}, eq 9.121.1).
This yields 
\bea
{\cal F}_0(n) & = & 1\,,\nonumber\\
{\cal F}_1(n) & = & 2n+1\,,\nonumber\\
{\cal F}_2(n) & = & 2n^2+2n+1\,,\nonumber\\ 
{\cal F}_3(n) & = & (2n+1)(2n^2+2n+3)/3\,,
\eea
and so on, after which the summation in \eq{momqm} becomes elementary.  
Below we give in \tab{momtab} (left part) a list of the first few 
moments, obtained in the TF approximation from \eq{momtf} and exactly 
from \eq{momqm}, expressed in terms of the Fermi energy $\lambda=M+1$. 
We see that the TF moments reproduce the exact ones up to terms of 
relative order $1/\lambda^2$ and lower, which are equivalent to terms 
of order $1/M^2$ and lower.

The same result can, with some more algebraic effort, be found for
higher dimensions by manual evaluation of \eq{momdqm}. For even $d$, the 
moments are more easily obtained from a Taylor expansion of the Laplace 
transforms $P(s)$ and $X(s)$ of the exact densities, given in appendix 
\ref{apposc}, in powers of $s$. For $d=2$, the simplicity of \eq{FG2}
allowed us to derive the recurrence relation
\be
C_{qm}^{(m+1)} = \left(\frac{m+1}{m+3}\right)\left[\,2\lambda\,
          C_{qm}^{(m)}+m^2 \,C_{qm}^{(m-1)}\,\right].\qquad\qquad (d=2)
\ee

In the rightmost column of \tab{momtab} we give the lowest 
quantum-mechanical moments for $d=2$, and in \tab{momtab2} we give
them for $d=3$ and $d=4$. Their leading terms reproduce in all cases 
the TF values \eq{momtf}. Note that the polynomials seen in these 
tables, translated into polynomials in $M$, appear exactly in the 
coefficients of the Taylor expansions of the densities for even $d$. 
This can be checked for $d=2$ and $d=4$ with the expansions given in 
the appendix \ref{apptay}.

This establishes the desired result that in the asymptotic limit, all
moments of the quantum-mechanical density $\rho(r,M)$ are identical with 
those of the TF density $\rho_{TF}(r,\lambda_M)$ up to relative terms of 
order $1/M^2$ or $1/\lambda_M^2$. By virtue of \eq{virial}, the same 
conclusion can be drawn for the kinetic energy density $\xi(r)$ since 
the proportionality constant in this equation does not depend on $M$ 
or $\lambda$. 

\Table{momtab}{16}{
\begin{tabular}{|c|l|l||l|}
\hline
$m$ & $C_{TF}^{(m)}\;(d=1)$ & $C_{qm}^{(m)}\;(d=1\,,\;\lambda=M+1)$
                            & $C_{qm}^{(m)}\;(d=2\,,\;\lambda=M+3/2)$ \\
\hline
0 & $2\lambda$  & $2\lambda$
                & $\lambda^2-\frac14$\\
1 & $\lambda^2$ & $\lambda^2 $  
                & $\frac23\,(\lambda^3-\frac14\lambda)$\\
2 & $\lambda^3$ & $\lambda^3+\frac12\lambda$  
                & $\frac23\,(\lambda^4+\frac12\lambda^2-\frac{3}{16})$\\
3 & $\frac54\,\lambda^4$ & $\frac54\,(\lambda^4+2\lambda^2)$  
             & $\frac45\,(\lambda^5+\frac52\lambda^3-\frac{11}{16}\lambda)$\\
4 & $\frac74\,\lambda^5$ & $\frac74\,(\lambda^5+5\lambda^3+\frac32\lambda)$  
             & $\frac{16}{15}\,(\lambda^6+\frac{25}{4}\lambda^4
                   +\frac{19}{16}\lambda^2-\frac{90}{128})$\\
5 & $\frac{21}{8}\,\lambda^6$ & $\frac{21}{8}\,(\lambda^6+10\lambda^4
                              +\frac{23}{2}\lambda^2)$  
                & $\frac{32}{21}\,(\lambda^7+\frac{49}{4}\lambda^5
                   +\frac{259}{16}\lambda^3-\frac{309}{64}\lambda)$\\
6 & $\frac{33}{8}\,\lambda^7$ & $\frac{33}{8}\,(\lambda^7+\frac{53}{2}\lambda^5
                              +49\lambda^3+\frac{45}{4}\lambda)$
                 & $\frac{16}{7}\,(\lambda^8+21\lambda^6+\frac{567}{8}\lambda^4
                   +\frac{89}{16}\lambda^2-\frac{1575}{256})$\\
\hline
\end{tabular}
}{~The lowest moments $C^{(m)}$ of the density of the 1-dimensional 
harmonic oscillator with $M+1$ filled shells, expressed in terms of 
the Fermi energy $\lambda=M+1$. Column 2: quantum-mechanical values 
from \eq{momqm}; column 3: TF values from \eq{momtf}. Column 4
contains the quantum-mechanical moments for the 2-dimensional
oscillator; the TF values (not shown explicitly) are given by the
leading term for each $m$.
}

\vspace*{-0.75cm}

\Table{momtab2}{16}{
\begin{tabular}{|c|l||l|}
\hline
$m$ & $C_{qm}^{(m)}\;(d=3\,,\;\lambda=M+2)$
    & $C_{qm}^{(m)}\;(d=4\,,\;\lambda=M+5/2)$ \\
\hline
0 & $\frac13\,(\lambda^3-\lambda)$
  & $\frac{1}{12}(\lambda^4-\frac52\lambda^2+\frac{9}{16})$\\
1 & $\frac14\,(\lambda^4-\lambda^2)$  
  & $\frac{1}{15}(\lambda^5-\frac52\lambda^3+\frac{9}{16}\lambda)$\\
2 & $\frac14\,(\lambda^5-\lambda)$  
  & $\frac{1}{15}(\lambda^6-\frac54\lambda^4-\frac{41}{16}\lambda^2
     +\frac{45}{64})$\\
3 & $\frac{7}{24}\,(\lambda^6+\frac52\lambda^4-\frac72\lambda^2)$  
  & $\frac{8}{105}(\lambda^7+\frac74\lambda^5-\frac{161}{16}\lambda^3
     +\frac{153}{64}\lambda)$\\
4 & $\frac38\,(\lambda^7+7\lambda^5-\frac72\lambda^3-\frac92\lambda)$  
  & $\frac{2}{21}(\lambda^8+7\lambda^6-\frac{133}{8}\lambda^4
     -\frac{177}{16}\lambda^2+\frac{945}{256})$\\
\hline
\end{tabular}
}{~Same as in column 4 of \tab{momtab}, but for $d=3$ and $d=4$.
}

\newpage

\section{Oscillating parts of the densities}
\label{secosc}

In this section we focus on the oscillating parts $\delta\rho(r)$ and 
$\delta\xi(r)$ of the densities as defined in \eq{oscden}. We first 
derive series expansions of these oscillating densities and then 
exhibit that their asymptotically leading terms are of order $1/M$ and 
$1/M^3$, respectively, relative to the leading-order terms of their
smooth parts. 

\subsection{Expansions of the exact oscillating parts}

We employ here a method proposed recently in \cite{zbsb}. In terms of 
the variable $t=2x=2r^2$, we define the Laplace transforms
\be
P(s) = \int_0^\infty e^{-st}\,\rho(t)\,\d t\,,\qquad
X(s) = \int_0^\infty e^{-st}\,\xi(t)\,\d t\,,
\label{lapden}
\ee
For even dimensions $d$ these can be evaluated analytically, as
shown in appendix \ref{apposc}, and the transforms of the smooth and 
oscillating parts can be readily identified. Taking the inverse Laplace 
transforms of the smooth parts yields exactly the results discussed in 
the previous section and summarized in \eq{TFas}, including the step
function which puts the TF densities identically to zero at and
beyond the classical turning point. The inverse Laplace transforms of 
the oscillating parts yield, after a suitable expansion for large $M$ 
which is shown in appendix \ref{apposc}, the following series for $d=2$:
\bea
\delta\rho^{(2)}(r)& = & \frac{(-1)^M}{2\pi}\left\{{\cal J}_0(z)
                         -\frac14\,x^2\,{\cal J}_2(z)
                         -\frac{\lambda}{9}\,x^3\,{\cal J}_3(z)
                         -\frac{1}{192}\,x^4\,{\cal J}_4(z)
                         +\dots\right\}, \label{drho2}\\
\delta\xi^{(2)}(r) & = & \frac{(-1)^M}{8\pi}\left\{-2\,x\,{\cal J}_1(z)
                         +\frac{1}{6}\,x^3\,{\cal J}_3(z)
                         +\frac{\lambda}{18}\,x^4\,{\cal J}_4(z)
                         +\frac{1}{480}\,x^5\,{\cal J}_5(z)
                         +\dots\right\}, \label{dxi2}
\eea
with $\lambda=M+3/2$ and
\be
z = 2\,\sqrt{2\lambda x}=2\,r\sqrt{2\lambda}\,.
\ee
Hereby we have defined the functions ${\cal J}_\mu(z)$ by 
\be
{\cal J}_\mu(z) = \Gamma(\mu+1)\left(\frac{2}{z}\right)^{\!\mu}\!
                  J_\mu(z) \qquad\qquad (\mu > -1)       \label{newbes}
\ee
in terms of the standard Bessel functions $J_\mu(z)$. Note that
the ${\cal J}_\mu(z)$ functions are eigenfunctions of the radial
Laplace operator, as discussed further below, and normalized such
that ${\cal J}_\mu(0)=1$.

For $d=4$ we find the results
\bea
\delta\rho^{(4)}(r)& = & \frac{(-1)^M}{4\pi^2}\left\{\lambda\,{\cal J}_1(z)
                         +\frac34\,x\,{\cal J}_1(z)
                         +\frac{\lambda}{6}\,x^2\,{\cal J}_2(z)
                         -\frac{1}{32}\,x^3\,{\cal J}_3(z)
                         -\frac{\lambda}{80}\,x^4\,{\cal J}_4(z)
                         +\dots\right\}\!, \label{drho4}\\
\delta\xi^{(4)}(r) & = & \frac{(-1)^M}{8\pi^2}\left\{{\cal J}_0(z)
                         -\frac{3}{4}\,x^2\,{\cal J}_2(z)
                         -\frac{\lambda}{9}\,x^3\,{\cal J}_3(z)
                         +\frac{1}{64}\,x^4\,{\cal J}_4(z)
                         +\frac{\lambda}{200}\,x^5\,{\cal J}_5(z)
                         +\dots\right\}\!,~~~~ \label{dxi4}
\eea
with $\lambda=M+5/2$. It is readily checked that in both cases, the 
above results satisfy the differential equation in \eq{difav}
connecting $\delta\xi(r)$ with $\delta\rho(r)$.

Using the known Taylor expansions of the Bessel functions, the above
results yield exactly the expansions of the oscillating terms 
given in appendix \ref{apptay} and discussed in the previous section. 
For the following it is important to note that the two leading powers 
in $M$ of each Taylor expansion coefficient are correctly reproduced 
by the leading-order Bessel functions in the above series; the 
higher-order terms just correct the terms of order $1/M^2$ and lower 
in each coefficient.

We were not able to find the same expansions of the oscillating
densities for odd $d$, since their Fourier transforms could not be 
obtained in a tractable analytical form. We have contented ourselves 
with the observation that the leading-order Bessel functions, given 
below for arbitrary $d$, are recovered from the Taylor expansions 
discussed in the previous section, again up to terms of relative 
order $1/M^2$ and lower in each Taylor coefficient.

The convergence of the above series is tested numerically in section 
\ref{secnum} below. They are asymptotic in nature and converge only 
for sufficiently small values of $x$, ie, sufficiently far from the 
classical turning points. At $x=0$, only their leading terms (ie, the 
first Bessel function in each series) contribute, for which we give 
some particularly nice closed forms in the following subsection.

\subsection{Asymptotically leading oscillating parts}

We now limit ourselves to the leading-order terms of the oscillating 
densities, which we will denote by the subscript ``as'' for 
``asymptotic'' and which are given by the first Bessel functions in 
each of the above exact expansions. For arbitrary $d$ and $M$ they can 
be written in the following forms:
\bea
\delta\rho_{\rm as}(r)\!& = &\!\frac{(-1)^M}{(2\pi)^{\mu+1}}\,
                            \left(\frac{2\lambda_M}{z}\right)^\mu\!J_\mu(z)\,,
                                                            \label{drhas}\\
\delta\xi_{\rm as}(r)\!& = &\!\frac{(-1)^M}{(2\pi)^{\mu+1}}\,\frac{(\mu+1)}{4}
                           \left(\frac{2\lambda_M}{z}\right)^{\mu-1} 
                           \!J_{\mu-1}(z)\,,                 \label{dxias}
\eea
with
\be
\mu = d/2-1\,, \qquad z = 2\,r\sqrt{2\lambda_M}\,, 
               \qquad \lambda_M = M+(d+1)/2\,.
\ee
Note that the above asymptotic forms fulfill the same differential 
equation as that given for the full oscillating parts in \eq{difav}. 
Comparing to the TF densities in \eq{tfden}, \eq{tfkin}, we see that 
$\delta\rho_{\rm as}(r)$ and $\delta\xi_{\rm as}(r)$ are of relative 
order $1/\lambda_M$ and $1/\lambda_M^3$, respectively. This confirms 
the numerical finding that even for moderate values of $M$, the 
densities $\xi(r)$ appear to be smooth, while the oscillations in 
$\rho(r)$ are clearly visible (see figures \ref{denf} and \ref{kinf}).

It is rather easy to verify that the asymptotic densities \eq{drhas},
\eq{dxias} are eigenfunctions of the Laplace operator for any $\mu$, 
and hence for any $d$, with eigenvalue $-8\lambda_M$:
\be
\Delta\,\delta\rho_{\rm as}(r) = -8\lambda_M\delta\rho_{\rm as}(r)\,,
\qquad 
\Delta\,\delta\xi_{\rm as}(r)  = -8\lambda_M\delta\xi_{\rm as}(r)\,.
\label{lapdrh}
\ee
Note, however, that this is not true for the full oscillating densities
given in \eq{drho2} -- \eq{dxi4} above. The equation  for 
$\delta\rho_{\rm as}(r)$ in \eq{lapdrh} can also be obtained from the 
exact differential equation \eq{difeq} derived in section \ref{secdifeq}. 
Taking the asymptotic limit of \eq{difeqosc}, retaining only the two 
leading powers in $\lambda_M$, the $\delta\xi(r)$ on the lhs and the 
potential term on the rhs can be neglected so that the first equation
in \eq{lapdrh} immediately follows.

Gleisberg {\it et al} \cite{glei} have studied the asymptotic expansion
in the case $d=1$ in which our result for $\delta\rho_{\rm as}$ becomes
\be
\delta\rho_{\rm as}^{(1)}(x) = \frac{(-1)^M}{\pi\sqrt{2\lambda}}\,
                               \cos(2x\sqrt{2\lambda})\,,\qquad
                               \lambda = M+1\,,
\ee
where $x$ here is the dimensionless coordinate. This result agrees with 
the oscillating part of their result (cf \cite{glei}, eqs 16 and A9), if 
we replace our Fermi energy $\lambda$ by $M+1/2$ which is that of the 
exact quantum system. (Note that these authors did not include the spin 
factor 2 in their density.) 

We emphasize that the asymptotic forms \eq{drhas}, \eq{dxias} give the
exact values of the oscillating densities at $r=0$, since all higher 
correction terms in the series \eq{drho2} -- \eq{dxi4} are zero at the
origin. As shown in the numerical investigations of section \ref{secnum}, 
they describe the quantum oscillations of the exact densities near the 
origin very precisely, where the densities are largest and perhaps most 
easily detected experimentally.

Next we derive the leading oscillating part of the kinetic energy 
densities $\tau(r)$ and $\tau_1(r)$ which are seen in \fig{kinf} to 
exhibit much stronger oscillations than $\xi(r)$. To derive the 
asymptotic forms of their oscillating parts, we start from 
\eq{taueqosc} and keep, as above, only the two leading-order terms in 
$\lambda_M$ on both sides of this equation. We then find the following 
simple relation
\be
\delta\tau_{\rm as}(r) = \lambda_M\,\delta\rho_{\rm as}(r)\,,
\label{taurho}
\ee
which will be tested numerically in section \ref{secnum}. Using 
the relations \eq{taurel} and \eq{lapdrh}, we finally obtain
\be
\delta\tau_{1,\rm as}(r) = -\lambda_M\,\delta\rho_{\rm as}(r)
                         = -\delta\tau_{\rm as}(r)\,,
\ee
which confirms explicitly the numerically known fact (cf \fig{kinf}) 
that the oscillations in $\tau_1(r)$ are opposite in phase to those 
in $\tau(r)$.

\subsection{Validity of the TF kinetic energy density functional}

In \cite{bvz}, the surprising observation was made that the TF
functional for the kinetic energy density
\be
\tau_{TF}[\rho_{TF}] = \frac{\hbar^2}{2m}\,\frac{4\pi d}{(d+2)}
                       \left[\frac{d}{4}\,\Gamma\!\left(\frac{d}{2}
                       \right)\right]^{\!2/d} \rho_{TF}^{\,1+2/d},
\label{tauofrho}
\ee
is fulfilled to a high degree of precision also for the exact
quantum-mechanical densities $\rho(r)$ and $\tau(r)$, except near 
the classical turning point and beyond it. Upon integrating, in 
the case $d=2$ one even obtains the exact kinetic energy.

Using the above asymptotic results it is easy to prove the validity 
of the TF functional \eq{tauofrho} at the level of the leading-order 
asymptotic terms. We define the asymptotic total densities by
\be
\rho_{\rm as}(r) = \rho_{TF}(r) + \delta\rho_{\rm as}\,,\qquad
\tau_{\rm as}(r) = \tau_{TF}(r) + \delta\tau_{\rm as}\,,
\label{asymden}
\ee
insert these into the TF relation \eq{tauofrho}, expand the power 
$[\rho(r)]^{\,1+2/d}$ up to first order in $\delta\rho_{\rm as}$, and 
neglect all terms of relative order $1/\lambda_M^2$ and lower. We 
then find that the TF functional relation \eq{tauofrho} is, indeed, 
fulfilled also by the total asymptotic densities \eq{asymden}
\be
\tau_{TF}[\rho_{\rm as}(r)] = \tau_{\rm as}(r) + {\cal O}(M^{-2})\,,
\label{taufunc}
\ee
which explains the results found numerically in \cite{bvz}. 

The result \eq{taufunc} is highly nontrivial. In principle, the TF relation 
$\tau_{TF}[\rho]$ holds only for systems with constant densities where 
the LDA is exact. For smoothly varying densities, gradient corrections 
are known to exist, which may be derived in the so-called extended TF 
model and have been successfully applied to various fermion systems (see 
eg \cite{book}, chapter 4). Nevertheless, our result \eq{taufunc} shows 
that the simple TF functional applies asymptotically for harmonically
trapped fermions also when including the quantum oscillations in both
$\rho(r)$ and $\tau(r)$.

\subsection{Numerical tests of the asymptotic relations}
\label{secnum}

We first test the asymptotic form \eq{drhas} of the oscillating part
of the density. In \fig{drhoas}, it is compared to the full oscillating
part $\delta\rho(r)$ defined by \eq{oscden} for $N=420$ particles in a 
2D trap. The cusp in the exact $\delta\rho(r)$ reflects the cut-off of 
the smooth TF-like density ${\bar\rho}(r)$ in \eq{avd2} at the classical 
turning point. We see that $\delta\rho_{as}$ in \eq{drhas} is correct 
only for small distances $r$ from the centre of the system and completely 
fails near the classical turning point. The situation improves somewhat 
when we include higher terms of the expansion discussed above. In 
\fig{drhoas2} we show the same comparison for $N=1722$ particles in 2D, 
with and without including the terms up to the order $x^8{\cal J}_8(z)$ 
in the Bessel function expansion \eq{drho2}. Whereas $\delta\rho_{as}$ 
is only sufficient up to $r\simeq 1$, the Bessel series truncated at 
order 8 becomes good up to $r\simeq 2.5$; the phases of the oscillations 
are correct even further out. But clearly, an infinite number of terms 
would have to be included before reaching an agreement in the region of 
the classical turning point.

\newpage

\Figurebb{drhoas}{42}{70}{795}{500}{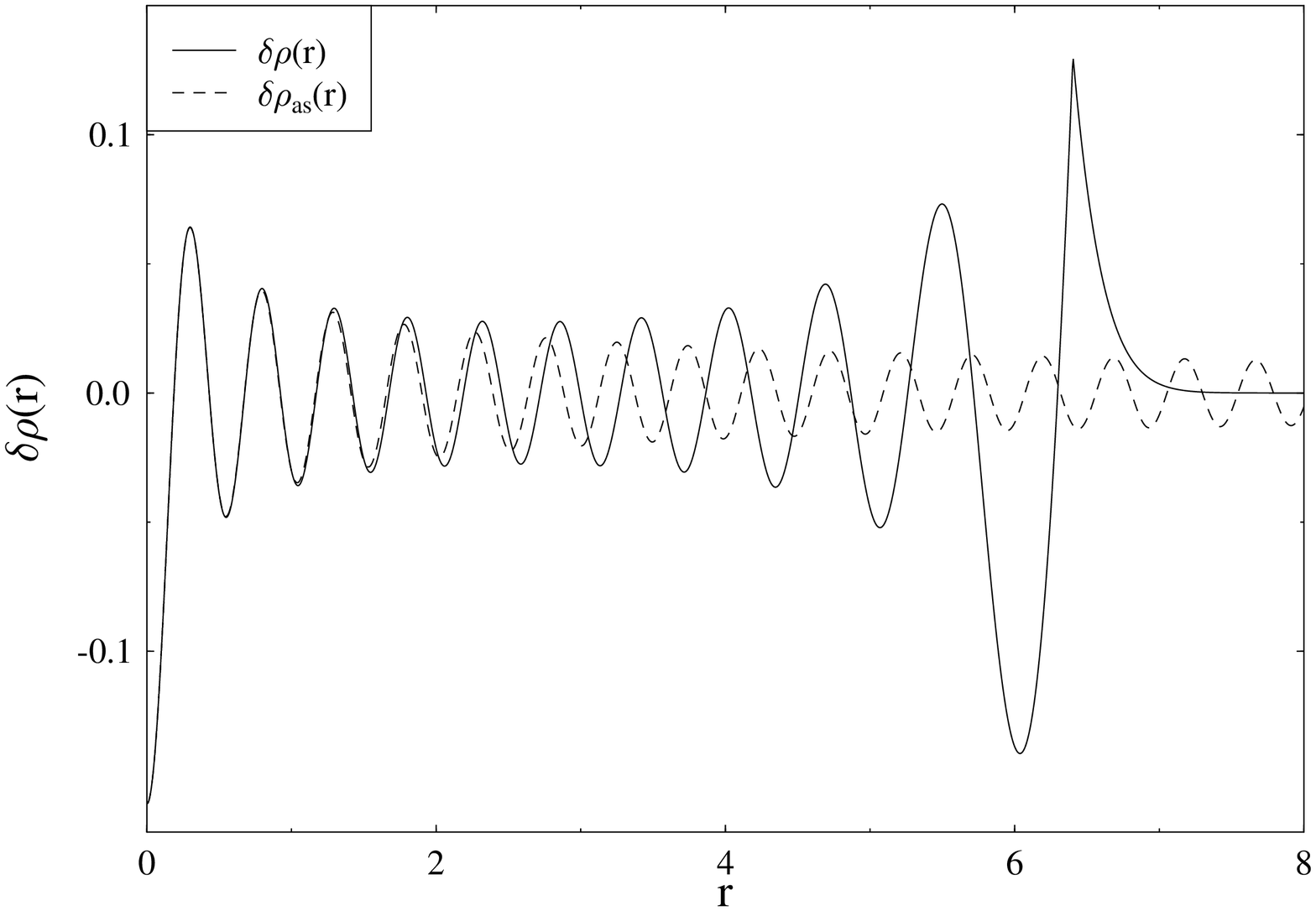}{8.5}{16}{
Oscillating part of the density for $N=420$ particles in a 2D harmonic 
trap ($M=20$). The solid line gives the full oscillating part as 
defined by \eq{oscden}, and the dashed line its asymptotic form given 
by \eq{drhas} for $\mu=0$, $\lambda=21.5$.
}

\vspace*{-0.75cm}

\Figurebb{drhoas2}{42}{70}{795}{500}{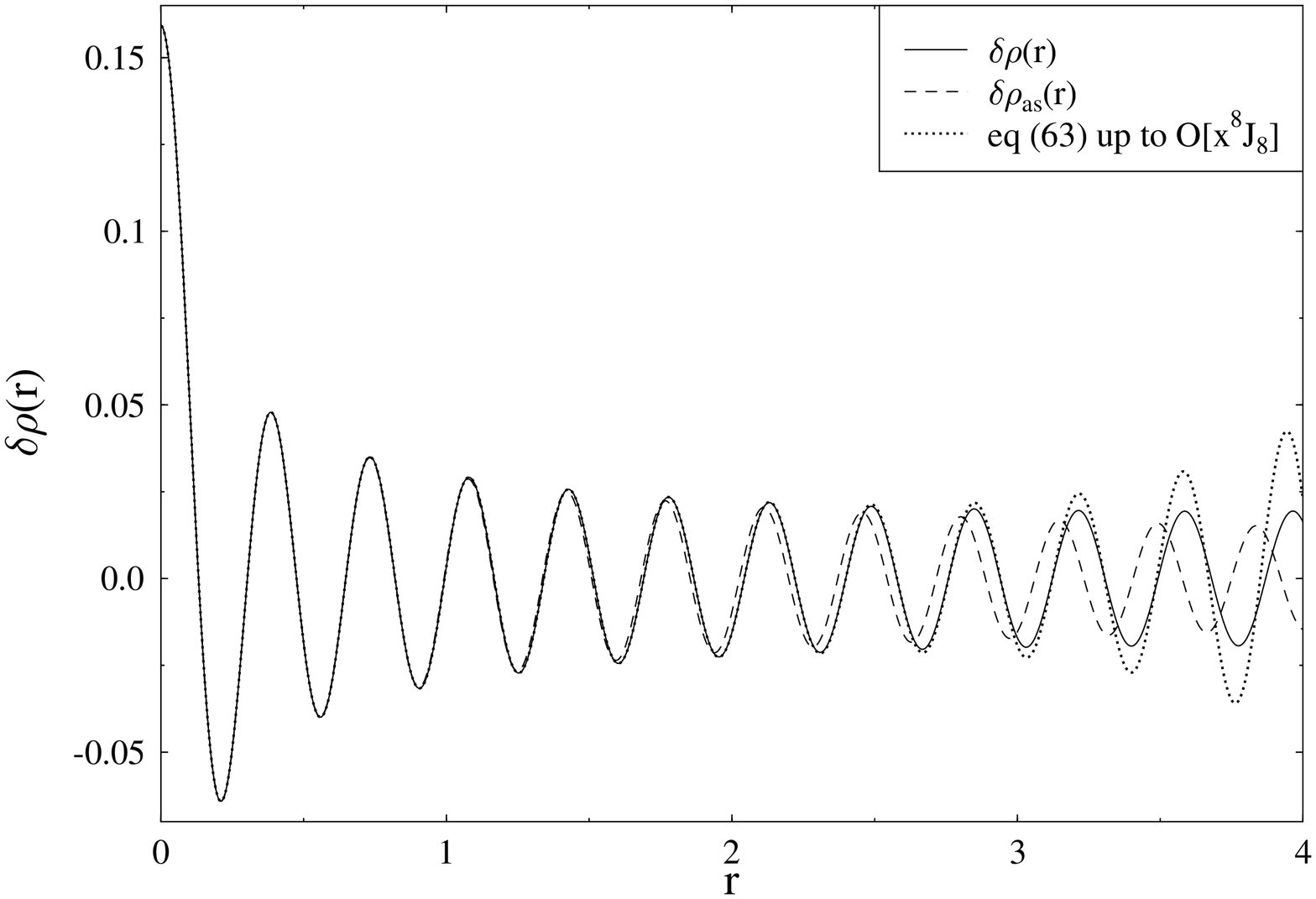}{8.5}{16}{
Same as \fig{drhoas}, but for $N=1722$ particles ($M=40$, $\lambda=41.5$). 
The dotted line gives the results obtained when including the
terms up to ${\cal O}[x^8{\cal J}_8(z)]$ in the expansion \eq{drho2}.
}

\vspace*{-0.5cm}

Whereas this result is not very encouraging, showing that the 
asymptotic oscillating densities \eq{drhas} are only useful far
away from the classical turning points, a much larger region of
validity is found for the relation \eq{taurho} between the 
asymptotic densities $\delta\tau_{as}(r)$ and $\delta\rho_{as}(r)$.
This is demonstrated in \fig{dtauas3}, where we have plotted the 
full quantities $\delta\tau(r)$ and $\lambda\,\delta\rho(r)$ (with 
$\lambda=62)$ for the $N=79422$ particles filling 61 shells in a
3D harmonic trap. The agreement is quite good until

\newpage

\Figurebb{dtauas3}{40}{30}{780}{260}{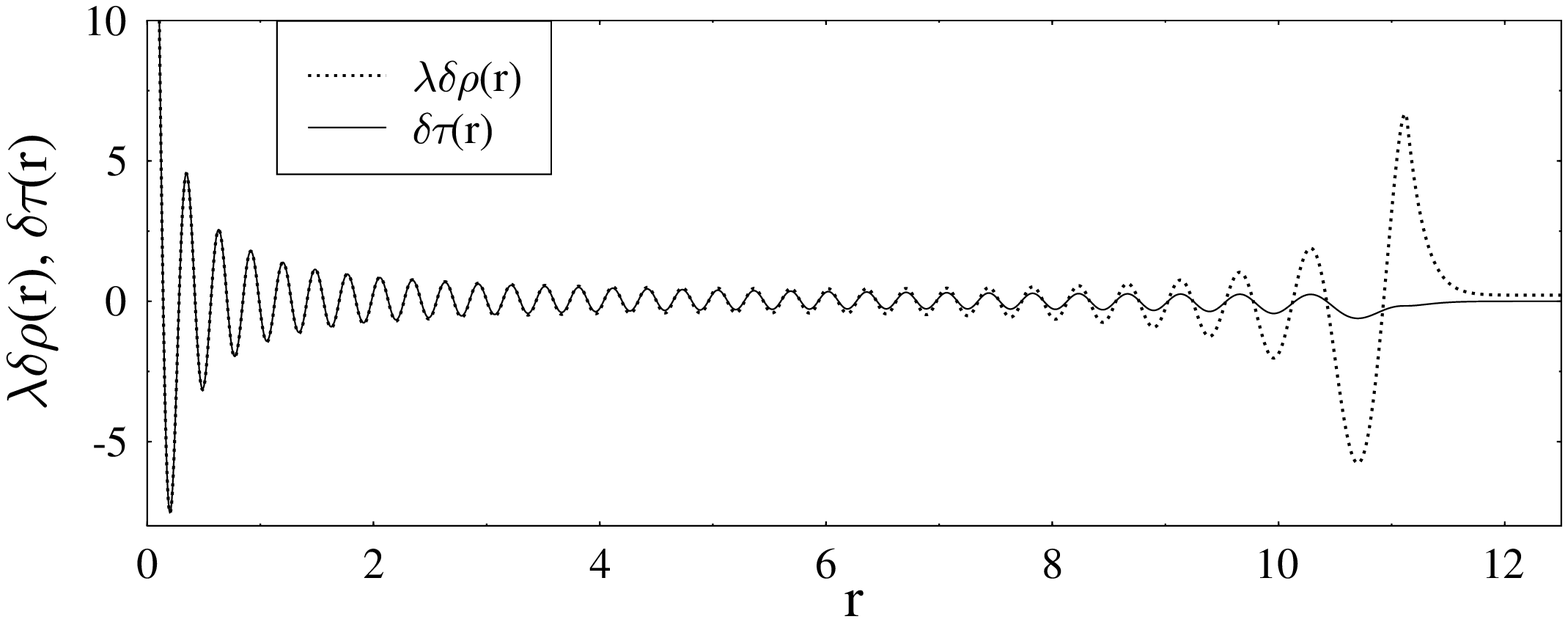}{5.2}{16}{
Oscillating parts of densities for $N=79422$ particles in a 3D 
harmonic trap ($M=60$, $\lambda=62$). The solid line gives 
$\delta\tau(r)$, and the dotted line the quantity 
$\lambda\delta\rho(r)$ according to the asymptotic relation 
\eq{taurho}.
}

\vspace*{-0.5cm}

\noindent
rather far away from 
the centre; only near the turning point the proportionality breaks down.
The same is shown in \fig{dtauas} for the $N=271502$ particles filling 
41 shells in a 4D harmonic trap (with $\lambda=42.5$). To set the scale, 
the density $\rho(r)$ and its TF limit are also shown in the figure. The 
particle number is so high here that the oscillations in $\rho(r)$ can 
only be recognized very close to the centre, which demonstrates once 
more the excellent approximation provided by the TF density. 

\Figurebb{dtauas}{42}{70}{780}{500}{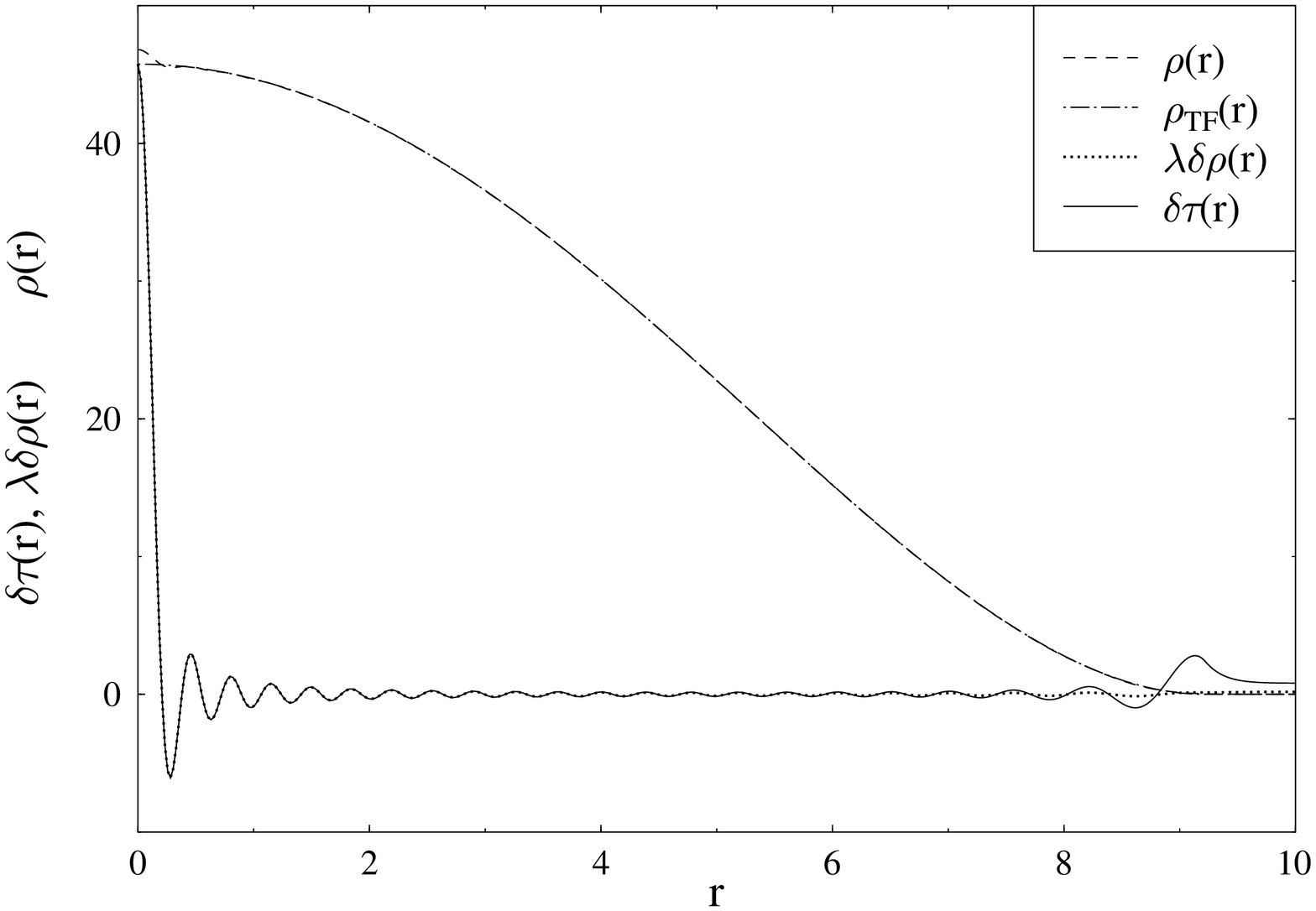}{9}{16}{
Same as \fig{dtauas}, but for $N=271502$ particles in a 4D harmonic 
trap ($M=40$) with $\lambda=42.5$. The dashed and dashed-dotted 
lines give the full density and its TF limit.
}

\vspace*{-0.5cm}

Clearly, an appreciable amount of error cancellation must take place 
on the way to the relation \eq{taurho} between the asymptotic 
oscillating densities. Its good validity is the basis of the 
performance of the TF kinetic energy functional $\tau_{TF}[\rho]$
including the quantum oscillations, as discussed above and
established formally through \eq{taufunc}.

%\newpage

\section{Summary}
\label{secsum}

We have studied the particle densities and several equivalent forms
of kinetic energy densities for harmonically trapped fermion gases
in arbitrary dimensions $d$. We established various exact differential
equations connecting these densities, and an equivalent 
integro-differential eigenvalue equation of Schr\"odinger type for the 
particle density. Some of these equations have been previously derived, 
but only in $d=1$ or 2 dimensions.
We have then studied in detail the asymptotic limits of the particle 
densities $\rho(r)$ and the kinetic energy densities $\xi(r)$ for
large particle numbers and Fermi energies $\lambda$. Using Taylor
expansions, moments, and Laplace transforms of these densities, we were
able to separate them uniquely into smooth and oscillating parts.
The two leading-order powers in $\lambda$ of the smooth densities 
${\bar\rho}(r)$, ${\bar\xi}(r)$ were shown analytically to go over into 
the well-known Thomas-Fermi (TF) densities, the remaining terms of 
order $1/\lambda^2$ and lower giving small corrections to the smooth 
parts. The Fermi energy hereby is given by $\lambda=M+(d+1)/2$, where 
$M$ is the quantum number of the highest occupied spherical shell.
The oscillating parts $\delta\rho(r)$, $\delta\xi(r)$ are characterized 
by an oscillating overall sign $(-1)^M$ that reflects the alternating 
parities of the successive shells. Their asymptotically leading terms 
could be given in simple closed expressions involving Bessel functions; 
they are eigenfunctions of the radial Laplace operator with eigenvalue
$-8\lambda$. They are of order $1/\lambda$ and $1/\lambda^3$ for 
$\delta\rho(r)$ and $\delta\xi(r)$, respectively, with respect to the
leading terms in the smooth densities ${\bar\rho}(r)$ and ${\bar\xi}(r)$.
The latter fact -- the suppression factor $1/\lambda^3$ -- explains the 
numerical experience that the kinetic energy densities $\xi(r)$ are 
smooth even for moderate particle numbers. For even dimensions $d$, we 
have derived a formally exact expansion of the full oscillating 
densities in terms of Bessel functions, which was, however, found to 
converge rather slowly and only sufficiently far away from the classical 
turning points. Still, these results are useful for expressing the
quantum oscillations near the centre of the system in closed analytical
forms. Using our exact differential equations which hold separately for 
the smooth and oscillating parts of the densities, we could also 
establish the numerical fact that the kinetic energy densities $\tau(r)$ 
and $\tau_1(r)$ are oscillating with opposite phases. An interesting 
proportionality between $\delta\rho(r)$ and $\delta\tau(r)$ was found 
to hold asymptotically everywhere except near the classical turning 
points. With our asymptotic relations, we could finally establish 
analytically the surprising recent observation \cite{bvz} that the TF 
functional $\tau_{TF}[\rho]$ holds also when the quantum-mechanical 
oscillations are included in both $\tau(r)$ and $\rho(r)$.

\bigskip

One of us (MB) acknowledges stimulating discussions with B van Zyl 
at early stages of these investigations. We are also grateful to 
R K Bhaduri for his continued interest. MB acknowledges the warm 
hospitality extended to him at the IMSC Chennai during a research 
visit, and financial support by the Deutsche Forschungsgemeinschaft.

\newpage

\begin{appendix}

\section{Evaluation or ${\cal R}(x)$}
\label{appdif}

In this appendix we give details on the evaluation of the quantity
${\cal R}(x)$ defined in \eq{residue} and used in deriving the 
differential equation \eq{difeq}. Here we prove the following
identity:
\begin{eqnarray}
{\cal R}(x) & = & \frac{2}{\pi^{\!d/2}}\sum_{\mu=0}^M (-1)^\mu
                  \left\{\left[(M+1-\mu)\,F_{M-\mu}^{(d)}-
                               G_{M-\mu}^{(d)}\right] 
                  L_\mu(2x)+\frac{(d-2)}{2}\,F_{M-\mu}^{(d)} L'_\mu(2x) 
                  \right\} e^{-x} \nonumber\\
       & \equiv & \frac{(d-2)}{d}\,\xi(x) - \frac{(d-2)}{4}\,\rho(x)\,.
\label{theorem}
\end{eqnarray}

\noindent 
{\it Proof:} Using the identity 
\be
L'_n(2x) = - \sum_{m=0}^{n-1} L_m(2x)\,,\qquad\qquad (n>0)
\ee
which may be derived using equations 8.971.2 and 8.974.3 of \cite{grry},
noting that $L'_0(2x) = 0$, and inverting the order of summation in the 
first line of \eq{theorem}, we get
\be
{\cal R}(x) = \frac{2}{\pi^{\!d/2}} \sum_{\mu=0}^M (-1)^{M-\mu}
              \left\{\!\left[(\mu+1)\,F_{\mu}^{(d)}-G_{\mu}^{(d)}\right]\! 
              L_{M-\mu}(2x)-\frac{(d-2)}{2}\,F_{\mu}^{(d)}\!
              \sum_{m=0}^{M-\mu-1}\!\!\! 
              L_m(2x) \!\right\} e^{-x}. 
\ee
(The sum over $m$ above must be put to zero when the upper limit of $m$ 
takes the value $-1$). We may now reorder the double sum over the last
term in such a way that the Laguerre polynomials can be taken out of the 
inner sum, leading to
\be
{\cal R}(x) = \frac{2}{\pi^{\!d/2}}\sum_{\mu=0}^M (-1)^{M-\mu}
              L_{M-\mu}(2x)\,H_{\mu}^{(d)}e^{-x},            \label{resimp}
\ee
with the definition
\be
H_{\mu}^{(d)} = (\mu+1)F_{\mu}^{(d)}-G_{\mu}^{(d)}-(-1)^{\mu}\,\frac{(d-2)}{2} 
                \sum_{\nu=0}^{\mu-1}(-1)^\nu F_{\nu}^{(d)}\,. \label{hcoeff}
\ee
For the sum over the $F_\nu^{(d)}$ on the rhs above, we can use the 
formulae \eq{Fodsum} given in section \ref{secdens}. Furthermore, from our
explicit results \eq{Fn} and \eq{Gn} one easily sees that
\begin{eqnarray}
(1+d/2)\,G_{2n+1}^{(d)} & = & (4n+2+d)\,F_{2n+1}^{(d)}\,,\nonumber\\
 (1+d/2)\,G_{2n}^{(d)}  & = & (4n+1+d/2)\,F_{2n}^{(d)}
                              +(d/2-1)\,F_{2n-1}^{(d)}\,.
\end{eqnarray}
Using these identities, it just takes some algebra to find the simple
form 
\be
H_{\mu}^{(d)} = \frac{(d-2)}{4}\left[G_{\mu}^{(d)}-F_{\mu}^{(d)}\right], 
\ee
valid for both even and odd $\mu$. Substituting this into \eq{resimp} 
we finally obtain the desired result \eq{theorem}.

\newpage

\section{Taylor expansions of exact densities}
\label{apptay}

We give here some Taylor expansions of the exact densities \eq{dden} 
and \eq{dkin} in powers of $x=r^2$. Their coefficients, defined in 
\eq{tayden}, cannot be given in closed form for any $d$ and $M$, but 
only separately for even and odd $M$. The coefficients $\rho^{(0)}(d,M)$ 
are simply the densities at $r=0$ given already in \eq{rhozero}. For the 
next three coefficients of the density $\rho(r)$, we find for even $M$:
\bea
\rho^{(1)}_{\rm even}(d,M) & = & -\rho^{(0)}_{\rm even}(d,M)\,,           
                                                           \label{rho1ev}\\
\rho^{(2)}_{\rm even}(d,M) & = & \frac{(2+d+4M)}{2\,(d+2)}\,
                             \rho^{(0)}_{\rm even}(d,M)\,, \label{rho2ev}\\
\rho^{(3)}_{\rm even}(d,M) & = & -\frac{(d^2+6\,d+8+12\,dM+16M+16M^2)}
             {6\,(d+4)(d+2)}\,\rho^{(0)}_{\rm even}(d,M)\,, \label{rho3ev}
\eea
and so on. For odd $M$ we find:
\bea
\rho^{(1)}_{\rm odd}(d,M) & = & \rho^{(0)}_{\rm odd}(d,M)\,, \label{rho1od}\\
\rho^{(2)}_{\rm odd}(d,M) & = & -\frac{(2+3\,d+4M)}{2\,(d+2)}\,
                                 \rho^{(0)}_{\rm odd}(d,M)\,,\label{rho2od}\\
\rho^{(3)}_{\rm odd}(d,M) & = & \frac{(5\,d^2+10\,d+8+20\,dM+16M+16M^2)}
               {6\,(d+4)(d+2)}\,\rho^{(0)}_{\rm odd}(d,M)\,, \label{rho3od}
\eea
and so forth. Similar expressions are obtained for the coefficients 
$\xi^{(m)}(d,M)$. We can then reconstruct separately the expansion 
coefficients of the smooth and oscillating parts defined in \eq{avden} 
and \eq{oscden}. We give here the explicit expansions, valid for both even 
and odd $M$, for even dimensions, where the results look more transparent. 
For $d=2$ we obtain
\bea
\rho_M^{(2)}(x) \!& = &\! \frac{1}{\pi}\,
                \left[\left(M+\frac32\right)-\frac12\,x\,\right]
                \;+\;\frac{(-1)^M}{2\pi}\left[1-2x\left(M+\frac32\right)
                +x^2\left(M^2+3M+2\right) \right.  \nonumber\\
                \!& &\! \left. \hspace{6cm} 
                -\,\frac29\,x^3\left(M^3+\frac92\,M^2+\frac{13}{2}\,M+3\right)
                + \dots \right]\!.                          \label{rho2exp}\\
\xi_M^{(2)}(x) \!& = &\! \frac{1}{2\pi}\,\left[\left(M^2+3M+2\right)
                     -x\left(M+\frac{3}{2}\right)+\frac14\,x^2\right]
                     \nonumber\\
    \!& - &\! \frac{(-1)^M}{4\pi}\left[\,x-x^2\left(M+\frac32\right)
              +\frac13\,x^3\left(M^2+3M+2\right)
        -\frac{1}{18}\,x^4\left(M^3+\frac92\,M^2+\frac{13}{2}M+3\right) 
         \right. \nonumber\\
   \! & &\! \left. \qquad\qquad 
        +\frac{1}{180}\,x^5\left(M^4+6M^3+14M^2+15M+6\right)
        + \dots \right].                                    \label{xi2exp}
\eea
For $d=4$ we get explicitly
\bea
\rho_M^{(4)}(x) \!& = &\! \frac{1}{4\pi^2}\,
                      \left[\left(M^2+5M+\frac{11}{2}\right)
                  -x\,\left(M+\frac52\right)+\frac14\,x^2\right]\nonumber\\
    \!& + &\! \frac{(-1)^M}{4\pi^2}\left[\left(M+\frac52\right)
          -x\,\left(M^2+5M+\frac{11}{2}\right)
          +\frac13\,x^2\left(M^3+\frac{15}{2}\,M^2+17M+\frac{45}{4}\right)
          \right.  \nonumber\\
    \!& &\! \left. \qquad\qquad 
        -\frac{1}{18}\,x^3\left(M^4+10M^3+35M^2+50M+24\right)
        + \dots \right],                                 \label{rho4exp}\\
\xi_M^{(4)}(x) \!& = &\! \frac{1}{6\pi^2}\,
                   \left[\left(M^3+\frac{15}{2}\,M^2+17M+\frac{45}{4}\right)
                     -\frac32\,x\left(M^2+5M+\frac{11}{2}\right)
                     +\frac34\,x^2\left(M+\frac52\right)-\frac18\,x^3\right]
                     \nonumber\\
    \!& + &\! \frac{(-1)^M}{8\pi^2}\left[1-2x\!\left(M+\frac52\right)
          +x^2\left(M^2+5M+\frac{11}{2}\right)
        -\frac29\,x^3\left(M^3+\frac{15}{2}\,M^2+17M+\frac{45}{4}\right) 
         \right. \nonumber\\
   \! & &\! \left. \qquad\qquad 
        +\frac{1}{36}\,x^4\left(M^4+10M^3+35M^2+50M+24\right)
        + \dots \right].                                  \label{xi4exp}
\eea
Note that the smooth and oscillating parts in the above expansions of
both densities are readily recognizable, the latter being multiplied 
by the overall sign $(-1)^M$.

\section{Asymptotic expansion of oscillating parts}
\label{apposc}

We start from the definition of the Laplace transforms \eq{lapden} of
the exact particle and kinetic energy densities \eq{dden} and \eq{dkin}. 
Using the integral (see \cite{grry}, equation 7.414.2)
\be
\int_0^\infty e^{-st}\,L_n(t)\,e^{-t/2}\,\d t 
             = \frac{(s-1/2)^n}{(s+1/2)^{n+1}}\,,
\ee
one obtains
\bea
P^{(d)}(s) & = & \frac{1}{\pi^{\!d/2}}\,2\sum_{\mu=0}^M
                 F_{M-\mu}^{(d)}\,\frac{(1/2-s)^\mu}{(1/2+s)^{\mu+1}}\,,
\nonumber\\
X^{(d)}(s) & = & \frac{1}{\pi^{\!d/2}}\,\frac{d}{2}\sum_{\mu=0}^M
                 G_{M-\mu}^{(d)}\,\frac{(1/2-s)^\mu}{(1/2+s)^{\mu+1}}\,. 
\label{lapdend}
\eea
We first discuss the simplest case, $d=2$, and reproduce the results of 
\cite{zbsb} in a slightly modified form. Here, as for any even dimension
$d$, the sums in \eq{lapdend} can be done analytically \cite{mapl}, with 
the results
\bea
P^{(2)}(s) & = & \frac{1}{\pi s}\left[(M+3/2)-\frac{1}{4s}\right]
               + \frac{(-1)^M}{4\pi s}\,
                 \frac{(1-1/2s)^{M+2}}{(1+1/2s)^{M+1}}\,,\label{F2s}\\
X^{(2)}(s) & = & \frac{1}{2\pi s}\left[(M^2+3\,M+2)-\frac{1}{2s}(M+3/2)
                                       +\frac{1}{16s^2}\right]
               + \frac{(-1)^{M+1}}{8\pi s^2}\,
                 \frac{(1-1/2s)^{M+2}}{(1+1/2s)^{M+1}}\,.~~~~\label{G2s}
\eea
These expressions fall into two parts which correspond exactly to the
Laplace transforms of the smooth and oscillating parts of the densities
defined in \eq{densep}, whereby we recognize the alternating sign 
$(-1)^M$ in front of the oscillating terms. The smooth parts give after 
Laplace inversion exactly the results \eq{avd2}. 

In the following we shall concentrate on the oscillating terms. We 
shall presently show that these terms yield oscillating functions of 
$r$ (or $x$) and derive their systematic asymptotic expansion. To this 
purpose we extract a common factor appearing in both functions above
\be
E(s) = \frac{(1-1/2s)^{M+2}}{(1+1/2s)^{M+1}}
     = \left(\frac{1-1/2s}{1+1/2s}\right)^{\lambda} 
       \left(1-\frac{1}{4s^2}\right)^{1/2}\,, \qquad \lambda=M+3/2\,.
\label{Es}
\ee
Using the substitution $s=\lambda/w$ we rewrite the first factor above
and evaluate its asymptotic expansion \cite{mapl} for 
$\lambda\rightarrow\infty$:
\be
\lim_{\lambda\to\infty} 
\left(\frac{1-w/2\lambda}{1+w/2\lambda}\right)^{\lambda}
= e^{-w}\left[1-\frac{w^3}{12\,\lambda^2}
  -\frac{1}{\lambda^4}\left(\frac{w^5}{80}-\frac{w^6}{288}\right)
  -\frac{1}{\lambda^6}\left(\frac{w^7}{448}-\frac{w^8}{960}
                            +\frac{w^9}{10368}\right)-\dots\right].
\ee
Next we re-insert $w=\lambda/s$ on the rhs above and perform a Taylor 
expansion of the second factor in \eq{Es} in powers of $1/2s$. Note 
that this expansion converges since we can always deform the contour 
of the inverse Laplace transform, which is a vertical line to the right 
of the imaginary axis in the complex $s$ plane, such that $|1/2s|<1$.
As a result, we obtain the following asymptotic expansion of $E(s)$:
\be
E(s) = e^{-\lambda/s}\left[1-\frac{1}{8s^2}-\frac{\lambda}{12s^3}
       -\frac{1}{128s^4}-\frac{\lambda}{480s^5}
       +\left(\frac{\lambda^2}{288}-\frac{1}{1024}\right)\frac{1}{s^6}
       -\frac{\lambda}{53670s^7}
       +\dots\right].
\ee
Together with the prefactors appearing in \eq{F2s} and \eq{G2s}, we
can now Laplace invert the resulting series term by term. Hereby we
use the general formula (see \cite{abro}, equation 29.3.80)
\be
{\cal L}_t^{-1}\left\{\frac{1}{s^{\mu+1}}e^{-\lambda/s}\right\}
= \left(\frac{t}{\lambda}\right)^{\mu/2}\!J_\mu(2\sqrt{t\lambda})\,, 
\ee
where the $J_\mu(z)$ are the cylindrical Bessel functions. Expressing
the latter in terms of the ${\cal J}_\mu(z)$ defined in \eq{newbes},
we arrive at the results \eq{drho2}, \eq{dxi2} given in section
\ref{secosc}. 

For the case $d=4$ we just give here the Fourier transforms of the 
oscillating parts 
\be
\delta P^{(4)}(s) = \frac{(-1)^M}{4\pi^2}\,\left[1-
                    \frac{(1-1/2s)^{M+4}}{(1+1/2s)^{M+1}}\right], \qquad
\delta X^{(4)}(s) = \frac{(-1)^M}{8\pi^2 s}\,
             \frac{(1-1/2s)^{M+4}}{(1+1/2s)^{M+1}}\,,
\ee
which can be again be clearly separated from their smooth parts that
yield the results \eq{avd4} after Laplace inversion. Proceeding exactly 
as above, but now using $\lambda=M+5/2$, we arrive at the series 
\eq{drho4}, \eq{dxi4} of the oscillating densities given in section 
\ref{secosc}.

Unfortunately, we could not find any tractable analytical expressions 
after the summations in \eq{lapdend} for odd dimensions $d$, which yield 
rather cumbersome combinations of hypergeometric series.

\end{appendix}


\begin{thebibliography}{31}

\setlength{\itemsep}{-0.25ex}

\bibitem{jin}  DeMarco B and Jin D S 1999 {\it Science} {\bf 285} 1703\\
               G\"orlitz A {\it et al.} 2001 {\it Phys.\ Rev.\ Lett.} 
               {\bf 87} 130402 

\bibitem{vig1} Vignolo P, Minguzzi A and Tosi M P 2000 
               {\it Phys.\ Rev.\ Lett.} {\bf 85} 2850

\bibitem{glei} Gleisberg F, Wonneberger W, Schl\"oder U and Zimmermann C
               2000 {\it Phys.\ Rev.} A {\bf 62} 063602

\bibitem{bvz}  Brack M and van Zyl B 2001 
               {\it Phys.\ Rev.\ Lett.} {\bf 86} 1574

\bibitem{mar1} Minguzzi A, March N H and Tosi M P 2001 {\it Eur.\ Phys.\ J.} 
               D {\bf 15} 315

\bibitem{mar2} Minguzzi A, March N H and Tosi M P 2001 {\it Phys.\ Lett.} A
               {\bf 281} 192

\bibitem{vig2} Vignolo P and Minguzzi A 2001 
               {\it J.\ Phys.\ B: At.\ Mol.\ Opt.\ Phys.} {\bf 34} 4653

\bibitem{homa} Howard I A and March N H 2001 {\it J.\ Phys.\ A: Math.\
               Gen.} {\bf 34} L491 

\bibitem{akde} Akdeniz Z, Vignolo P, Minguzzi A and Tosi M P 2002
               e-print cond-mat/0205480

\bibitem{zbsb} van Zyl B, Bhaduri R K, Suzuki A and Brack M, 2002
               {\it Phys.\ Rev.} A submitted; e-print cond-mat/0209460

\bibitem{rkb1} Bhaduri R K and Zaifman L F 1979 {\it Can.\ J.\ Phys.} 
               {\bf 57} 1990\\
               Guet C and Brack M 1980 {\it Z.\ Phys.} A {\bf 297} 247

\bibitem{mapl} we have used the mathematical software package 
               Maple$^{\copyright}$

\bibitem{abro} Abramowitz M and Stegun I A 1970 {\it Handbook of Mathematical
               Functions} (Dover Publications, 9th printing, New York)

\bibitem{lama} Lawes G P and March N H 1979, {\it J.\ Chem.\ Phys.}
               {\bf 71} 1007

\bibitem{grry} Gradshteyn I S and Ryzhik I M 1994 {\it Table of
               Integrals, Series, and Products} (Academic Press, New
               York, 5th edition)

\bibitem{book} Brack M and Bhaduri R K 1997 Semiclassical Physics 
               {\it Frontiers in Physics} vol 96 (Reading, MA: 
               Addison-Wesley) Revised paperback edition to appear in 
               January 2003 (Boulder, CO: Westview Press)

\end{thebibliography}
\end{document}